\documentclass[sigconf, nonacm]{acmart}
\AtBeginDocument{%
  }

\setcopyright{none}
\renewcommand\footnotetextcopyrightpermission[1]{}

\newcommand{\contextquote}[3][]{{{\emph{``{#3}''}\,---\,[#2]}}}
\usepackage{tikz}
\usepackage[tikz,usetwoside=true]{mdframed}
\usepackage{amsmath}

\newcommand{\No}{\protect\textsc{No-AI}}
\newcommand{\Free}{\protect\textsc{Free-AI}}
\newcommand{\Paid}{\protect\textsc{Paid-AI}}

\usetikzlibrary{calc,arrows}
\tikzset{
    summary arrow/.style={%
        line width=1pt,
        draw=gray!40,
        rounded corners=1ex,
    },
    summary head/.style={
        fill=white,
        font=\bfseries\sffamily,
        text=gray!80,
        anchor=base west,
    },
}
\mdfdefinestyle{summary}{%
    singleextra={%
        \path let \p1=(P), \p2=(O) in (\x2,\y1) coordinate (Q);
        \path let \p1=(P), \p2=(O) in (\x1,\y2) coordinate (R);
        \path let \p1=(Q), \p2=(O) in (\x1,{(\y1-\y2)/2}) coordinate (M);
        \path [summary arrow] (R) -| (M) |- (P);
        \node [summary head] at ($(Q)+(1ex,-1.5pt)$) {\frametitle};
    },
    firstextra={%
        \path let \p1=(P), \p2=(O) in (\x2,\y1) coordinate (Q);
        \path [summary arrow,latex-] (O) |- (P);
        \node [summary head] at ($(Q)+(1ex,-1.5pt)$) {\frametitle};
    },
    secondextra={%
        \path let \p1=(P), \p2=(O) in (\x2,\y1) coordinate (Q);
        \path let \p1=(P), \p2=(O) in (\x1,\y2) coordinate (R);
        \path [summary arrow] (R) -| (Q);
    },
    middleextra={%
        \path let \p1=(P), \p2=(O) in (\x2,\y1) coordinate (Q);
        \path [summary arrow] (O) -- (Q);
    },
    middlelinewidth=1.5em,middlelinecolor=white,
    hidealllines=true,topline=true,
    innertopmargin=0pt,
    innerbottommargin=1em,
    innerrightmargin=0pt,
    innerleftmargin=1.5ex,
    skipabove=0.5\baselineskip,
    skipbelow=0.3\baselineskip,
}

\newenvironment{summary}[1]{
\begingroup
\def\frametitle{\small #1}
\begin{mdframed}[style=summary]
}{
\end{mdframed}
\endgroup
}

\usepackage{booktabs}
\usepackage{color, colortbl}
\definecolor{LightGray}{gray}{0.95}
\usepackage[inkscapelatex=false]{svg}
\usepackage{comment}
\usepackage{hyperref}
\usepackage{enumitem}
\setlist{noitemsep}
\usepackage{xurl}
\usepackage{tabularx}

\setcopyright{none}
\renewcommand\footnotetextcopyrightpermission[1]{}

\begin{document}

\title[The Impact of AI-Assisted Development on Software Security: A Study of Gemini and Developer Experience]{The Impact of AI-Assisted Development on Software Security:\\ A Study of Gemini and Developer Experience}

\author{Nadine Jost}
\email{nadine.jost@rub.de}
\affiliation{
  \institution{Ruhr University Bochum}
  \city{Bochum}
  \country{Germany}
}

\author{Benjamin Berens}
\email{benjamin.berens@kit.edu}
\affiliation{
  \institution{Karlsruhe Institute of Technology}
  \city{Karlsruhe}
  \country{Germany}}

\author{Manuel Karl}
\email{m.karl@tu-braunschweig.de}
\affiliation{
  \institution{TU Braunschweig}
  \city{Braunschweig}
  \country{Germany}}

\author{Stefan Albert Horstmann}
\email{stefan.horstmann@uni-koeln.de}
\affiliation{%
  \institution{University of Cologne}
  \city{Cologne}
  \country{Germany}}

\author{Martin Johns}
\email{m.johns@tu-braunschweig.de}
\affiliation{%
  \institution{TU Braunschweig}
  \city{Braunschweig}
  \country{Germany}}

\author{Alena Naiakshina}
\email{alena.naiakshina@uni-koeln.de}
\affiliation{%
  \institution{University of Cologne}
  \city{Cologne}
  \country{Germany}}

\renewcommand{\shortauthors}{Jost et al.}

\begin{abstract}
The ongoing shortage of skilled developers, particularly in security-critical software development, has led organizations to increasingly adopt AI-powered development tools to boost productivity and reduce reliance on limited human expertise. These tools, often based on large language models, aim to automate routine tasks and make secure software development more accessible and efficient. However, it remains unclear how developers' general programming and security-specific experience, and the type of AI tool used (free vs. paid) affect the security of the resulting software.
Therefore, we conducted a quantitative programming study with software developers (n=159) exploring the impact of Google's AI tool Gemini on code security. Participants were assigned a security-related programming task using either no AI tools, the free version, or the paid version of Gemini. 
While we did not observe significant differences between using Gemini in terms of secure software development, programming experience significantly improved code security and cannot be fully substituted by Gemini.
\end{abstract}

\keywords{Developer Security Study, Gemini, LLMs, Artificial Intelligence, AI Assistants}

\maketitle

\section{Introduction}
\textit{``Imagine hiring a brilliant junior developer who arrives every morning with no memory of yesterday’s lessons, infinite enthusiasm but no judgment, and the ability to produce 1,000 lines of code while you grab coffee. Now imagine working with them on a year‐long project''}~\cite{hassan2026agentic}.
The software industry is facing a persistent and growing shortage of skilled developers, particularly in security-critical domains~\cite{mattioli2024foresight}. This problem is driven by several interrelated factors, including the rapid pace of technological advancement, insufficient adaptation of educational systems, and increasing demand across various industries~\cite{CanGener5:online}. Despite the rise of short courses and certifications, the supply of developers with strong software and security skills continues to fall short of meeting the needs of companies undergoing digital transformation. To address this gap, organizations are increasingly turning to AI-powered development tools, including both free and commercial offerings, as a means of improving productivity and reducing reliance on scarce human expertise (e.g.,~\cite{Chandra:online, SurveyTh78:online, Howsoftw68:online}). These tools, often based on large language models (LLMs), promise to automate routine coding tasks, accelerate development workflows, and make software engineering more accessible~\cite{2024Stac15:online, Surveyre91:online}.
A survey by GitHub found that 92\% of U.S.-based developers are now employing AI tools in both professional and personal capacities~\cite{Surveyre91:online}.
While past research investigating software security~\cite{pearce_asleep_2021, sandoval_lost_2022, perry_users_2022, asare_user-centered_2024}, revealed that around 40\% of the AI-generated code contained vulnerabilities~\cite{pearce_asleep_2021}, it remains unclear how developers' general programming and security-specific experience affect the security of the resulting software. For example, a recent outage prompted Amazon to involve senior engineers to address issues arising from ``Gen-AI-assisted'' changes~\cite{morales2026amazon_genai_changes}.

ChatGPT~\cite{ChatGPTO21:online}, GitHub Copilot~\cite{GitHubCo85:online}, and Gemini~\cite{Gemini57:online} are the most popular and well-known examples of AI tools among developers~\cite{2024Stac15:online}.
While the impact of Copilot and ChatGPT on code security has been studied~\cite{hamer_just_2024, oh_poisoned_2023, siddiq_empirical_2022, pearce_asleep_2021, sandoval_lost_2022, perry_users_2022, asare_user-centered_2024}, the influence of Google's AI Gemini, previously known as Bard~\cite{GoogleBa3:online}, remains underexplored. Therefore, we focused on Gemini as a case study and conducted an online study.
Launched in 2024, Gemini’s paid version, Gemini Advanced, is based on a different underlying model than the free version and introduced enhanced capabilities, including multimodal processing and advanced reasoning~\cite{GoogleGe4:online}.
In a study on consumers’ expectations and behavior regarding apps~\cite{bamberger2020can}, users reported differences in security and privacy between free and paid versions. While free versions were more likely to be installed, participants placed greater trust in paid versions to comply with security and privacy standards. However, it remains unclear whether similar effects apply to software developers. Thus, we tested both the free and paid versions of Gemini.

We conducted an online study with 159 developers recruited via Upwork~\cite{UpworkTh45:online} and assigned participants to either the paid or free version of Gemini, or to a control group without access to AI tools. We asked them to complete a programming task that included user authentication and functionality for managing a list of websites associated with a user while considering secure implementation.  After task completion, we evaluated participants’ submissions for the five common vulnerabilities: Cross-Site Scripting (XSS), Cross-Site Request Forgery (CSRF), Improper Input Validation, SQL Injection, and Cryptographic Failures.

With this work, we examined the impact of varying levels of developer expertise and the use of paid versus free AI coding tools on software security, as well as how trust in these tools influences developers' behavior:

\begin{itemize}
    \item \textbf{RQ1}: How does developers’ programming experience affect the security of the developed software?
    \item \textbf{RQ2}: How does developers’ security experience affect the security of the developed software?
    \item \textbf{RQ3}: How does the use of paid versus free AI-assisted development tools such as Gemini impact security?
    \item \textbf{RQ4}: Do developers exhibit differing levels of trust in Gemini's free and paid versions?
\end{itemize}

We found that participants' programming experience significantly improved the security of code submissions, while no significant differences were observed in the security of code written with the assistance of the free version of Gemini, the paid version, or no assistance. Further, no significant security benefit or difference in user trust was found between the paid and free versions.
Our results suggest that while Gemini can serve as valuable supplementary aid, it cannot fully substitute programming experience.

\section{Related Work}
This section outlines studies on AI tool code generation and security developer studies.

\subsection{AI Assistance Tool Studies}
Research on transformer-based LLMs like Codex, AlphaCode, and GPT-4 assessed their performance in code generation~\cite{li_competition-level_2022, sobania_choose_2022, hou_systematic_2024, du_evaluating_2024} and developed benchmarks like HumanEval~\cite{chen_evaluating_2021} and MBPP~\cite{austin_program_2021}, highlighting their potential to enhance software development. Developers value AI tools for boosting productivity~\cite{StackOve10:online, liang_large-scale_2024}, with 92\% of U.S. developers using AI tools for programming professionally and personally~\cite{Surveyre91:online}.

Pearce et al.~\cite{pearce_examining_2023} found that LLMs could fix 100\% of security bugs in carefully constructed prompts. Siddiq et al.~\cite{siddiq_empirical_2022} identified 265 code smell types and 44 security smells across 3 LLM training sets, including 18 code smells and 2 security smells in GitHub Copilot's suggestions. He and Vechev~\cite{he_large_2023} proposed SVEN, a learning-based approach that improved secure code generation from a CodeGen LM from 59.1\% to 92.3\%, while maintaining functional correctness.

Pearce et al.~\cite{pearce_asleep_2021} analyzed 1,689 GitHub Copilot-generated programs for vulnerabilities relevant to the ``2021 CWE Top 25 Most Dangerous Software Weaknesses''~\cite{CWE2023C20:online}, finding issues in about 40\% of the programs. Mousavi et al.~\cite{mousavi_investigation_2024} investigated 5 Application Programming Interfaces (APIs) and found that ChatGPT generated code with API misuses in around 70\% of the cases in a set of 48 programming tasks and revealed 20 distinct misuse types.
Whether developers would prompt AI tools similarly or adopt their suggestions remains uncertain.

Sandoval et al.~\cite{sandoval_lost_2022} studied 58 students using OpenAI’s code-cushman-001 to implement a shopping list structure in C, manually analyzing the code for CWEs. Students with AI access produced security bugs at a rate no higher than 10\% compared to those without, suggesting that LLMs did not introduce new security risks.

Perry et al.~\cite{perry_users_2022} conducted a study with 47 students and professionals to explore interactions with an AI assistant based on OpenAI’s codex-davinci-002 model for security-related tasks in Python, JavaScript, and C. 
In contrast to~\cite{sandoval_lost_2022}, they showed that code written with access to AI was less secure compared to code written without access. Interestingly, participants with security experience were less likely to trust and replicate AI outputs than those without security experience.

Asare et al.~\cite{asare_user-centered_2024} evaluated GitHub Copilot’s security performance in a user study with 25 students and professional developers, comparing code written with and without Copilot assistance in 2 C tasks: implementing a user sign-in and a transaction fulfillment. They found that access to Copilot accompanied more secure solutions when tackling more complex problems, and for easier problems, no effect was observed. 
Similarly to~\cite{sandoval_lost_2022}, no disproportionate impact on particular vulnerabilities was found, highlighting the potential security benefits of using Copilot. 

Oh et al.~\cite{oh_poisoned_2023} conducted 2 studies on developer responses to insecure AI code suggestions. An online survey of 238 developers found they frequently use these tools but may overlook poisoning risks. In a lab experiment with 30 professionals, developers using a poisoned tool were more likely to write insecure code and participants trusted code completion tools significantly more than code generation tools.

A further study on consumer expectations and app usage behavior~\cite{bamberger2020can} found that users perceived notable security and privacy differences between free and paid versions of apps. Similarly, in healthcare research~\cite{musheyev2024readability}, paid chatbots were found to provide more readable responses.
Compared to previous work, we focused our research on the security implications of using Gemini for software development and how paid and free AI versions affect code security and trust.

\subsection{Security Developer Studies}
Acar et al.~\cite{acar_you_2016} examined the impact of information resources on code security by surveying 295 Google Play developers and conducting a lab study with 54 Android developers. 
The findings showed developers using Stack Overflow produced less secure code than those using official documentation or books. Nadi et al.~\cite{nadi_jumping_2016} found that developers struggle with low-level Java cryptography APIs despite confidence in concept selection, highlighting a need for more task-based support.

Further, Acar et al.~\cite{acar_security_2017} conducted an online programming study with 307 GitHub users, who completed various security-related tasks, including credential storage, encryption, and writing a URL shortener service. They found differences in security and functionality based on participants' self-reported years of experience. 

Naiakshina et al.~\cite{naiakshina_why_2017, naiakshina_deception_2018, naiakshina_if_2019, naiakshina_conducting_2020} investigated developers' security behavior with secure password storage. 
In a lab study~\cite{naiakshina_why_2017}, 20 students were tasked with storing passwords; half were told the study was about API usability, while the other half were directly instructed to ensure secure password storage.
The authors noted that understanding security concepts did not always lead to secure task solutions.
In a follow-up study, Naiakshina et al.~\cite{naiakshina_deception_2018} found a positive effect of copy/paste on security. 
This study also analyzed the effect of programming experience on the security scores, showing that years of Java experience had no significant effect on security. 
In contrast to the previous work, we wanted to understand if AI tools can substitute programming and security experience.

\section{Methodology}
We designed an experiment to examine developers' programming and security experience and the effect of Gemini's free and paid versions on code security score.
Based on our power analysis (see Section~\ref{sec:PowerAnalysis}), we required a large sample of a hard-to-reach population (n = 159). We therefore conducted a remote field study with a large, international sample in their familiar environment, recruited via Upwork~\cite{UpworkTh45:online}, and compared 3 groups solving security-related tasks: 

\begin{itemize}
    \item \textbf{\No{}}: A control group that did not use any AI.
    \item \textbf{\Free{}}: A group that used only Gemini's free version.
    \item \textbf{\Paid{}}: A group that used only Gemini's paid version.
\end{itemize}

Following best practices that recommend instructing participants not to use LLMs when they are not permitted~\cite{prolific_llm_detection}, participants in \No{} were instructed not to use AI.
By contrast, participants in \Free{} were asked to use the free version of Gemini with their own Google accounts, while participants in \Paid{} were provided with the login credentials for a Google account with a paid Google One subscription including access to Gemini Advanced. 
After finishing the implementation task, participants submitted the code via GitHub Classroom and completed a survey. The survey included questions on demographics, challenges faced during implementation, support and difficulties with Gemini, and trust in AI. We additionally used standardized scales to measure the workload using the NASA-TLX~\cite{hart1988development}, usability of Gemini using the System Usability Scale (SUS)~\cite{brooke1996sus}, and participants' security self-efficacy using the Secure Software Development Self-Efficacy Scale (SSD-SES)~\cite{votipka_building_2020}.

\subsection{Programming Task}
Due to the constantly growing cloud business, web applications are becoming increasingly important compared to conventional local on-premise solutions~\cite{cloud_revenue}.
Therefore, we designed a web-based programming task similar to Linktree~\cite{Linkinbi74:online} for our study.
Linktree is a tool that lets users create a single link to house multiple links to their social media profiles, websites, and other online resources.
We chose Python as the programming language due to its first-place ranking on the TIOBE Index~\cite{TIOBEInd0:online} and its position as the most popular language on GitHub~\cite{GithubLa92:online}. Additionally, we used the Flask framework because it was the most popular Python-based web framework in 2024~\cite{2024Stac15:online}.

For our study, we examined the top 10 of the 25 most dangerous CWEs~\cite{CWE2023C20:online} to identify relevant vulnerabilities for the study task and analyze the submissions' security.
We excluded Out-of-bounds Write (CWE-787), Use After Free (CWE-416), and Out-of-bounds Read (CWE-125) due to their specific applicability to low-level programming languages like C or C++. Further, we excluded OS Command Injection (CWE-78) because its characteristics depend on the underlying operating system.
Finally, we excluded Path Traversal (CWE-22) and Unrestricted Upload of File with Dangerous Type (CWE-434), as file uploads and file system interactions were not aligned with our programming task.
Thus, we selected the remaining 4 vulnerabilities:\\
\textbf{Cross-Site Scripting (XSS):} User input is not properly sanitized before being included in web pages, allowing it to be executed as code and potentially leading to malicious code execution (CWE-79)~\cite{CWECWE792:online}.\\
\textbf{Cross-Site Request Forgery (CSRF):} The web application cannot fully verify if a valid request was intentionally submitted by the user (CWE-352)~\cite{CWECWE3536:online}.\\
\textbf{Improper Input Validation:} Input is received without proper validation to ensure it meets the required properties for safe and correct processing (CWE-20)~\cite{CWECWE2047:online}.\\
\textbf{SQL-Injection:} User-controller input is used in an SQL command without properly neutralizing special elements, allowing user inputs to be interpreted as SQL rather than as data (CWE-89)~\cite{CWECWE8999:online}.

In addition, we %
selected the OWASP Top Ten category \textbf{Cryptographic Failures}~\cite{A02Crypt35:online}, as authentication is essential for any web application that applies user accounts. Yet, password-based authentication is the most commonly used authentication method in sensitive areas such as finance and medical care~\cite{Authenti35:online, Authenti52:online}.
Further, a dataset of 9,948,575,739 unique plaintext passwords, uncovered by researchers in 2024~\cite{cybernews}, and a ruling against Meta by the Irish Data Protection Commission for storing passwords in plaintext~\cite{meta_judgment} highlight ongoing issues with inadequate password storage practices.

With our final set of 5 vulnerabilities, we designed the programming task that included user authentication and functionality for managing a list of websites associated with a user. The task was divided into 4 subtasks: implementing 1) a user registration, 2) a user login, 3) a function for adding websites, and 4) a function for deleting websites.
All participants were asked to pay attention to security while completing the task. Participants could use their usual IDE and consult any resources to solve the problem, except for restrictions on using Gemini.
A detailed description of the task can be found in the Appendix~\ref{task}.

\subsection{Pilot Study} 
We conducted a series of pilot studies to test our task design. 
First, the study was piloted with 2 student assistants. The first pilot participant worked approximately 2.5 hours on the task. Following the examples set in related AI studies~\cite{perry_users_2022, asare_user-centered_2024}, we aimed for a 2-hour duration for our study task. Thus, we removed one of the initial 5 subtasks from the code. This subtask was implementing the public profile view of users, which had the lowest security impact.
In the following, we piloted our 3 groups with 1 participant per group from Upwork. These pilot participants worked 108 minutes on average. 
One participant canceled their participation after knowing they were assigned to the \No{} group. 
This led us to add deception to the study design and not to tell participants in the actual study that 3 groups and AI tools were being investigated. 
They were told the study was about developers' programming behavior with a Flask web application.

\subsection{Power Analysis}
\label{sec:PowerAnalysis}
Prior to conducting the experiment, a power analysis was performed to determine the necessary sample size. We considered a medium effect size (f = 0.25)~\cite{cohen_statistical_1988}, a significance level of $\alpha = 0.05$, a power of 0.80, a numerator degree of freedom of 2, and 3 groups in the experimental design. The power analysis indicated that a sample size of 159 participants would provide sufficient statistical power to detect statistical significance. Consequently, a sample size of 159 participants (53 per group) was determined for the study to ensure the robustness and reliability of the findings.

\subsection{Participants}

We recruited freelance software developers on Upwork~\cite{UpworkTh45:online}, following the recommendations of prior work on developer recruitment best practices~\cite{kaur_where_2022}. 
The job post for this study can be found in the Appendix~\ref{jobpost}. Eligible participants were at least 18 years old and were proficient in English. Further, participants were required to have Python skills listed in their Upwork profiles. 
The Upwork team supported us during recruitment and invited 2138 people to our job post. Of these, 542 answered the invitation. The Upwork team screened the interested participants directly based on the skills advertised in their Upwork profiles. The researchers double-checked the screening criteria, and 291 eligible participants were contacted. 102 participants did not respond or canceled.
Finally, 189 participants were recruited and assigned to one of our 3 groups. 
All participants were compensated with \$60 via Upwork. 

We asked participants to submit their solutions within 2 days to reflect that freelancers might work on parallel projects. We chose this period to prevent the paid AI accounts from being blocked for other participants for a longer period of time, as these accounts could only be used by one participant at a time. All participants were asked in the follow-up survey what resources they used to complete the task.
From the 189 participants who completed the study, 7 were excluded in \No{} for using AI tools despite being told not to, 14 participants were excluded in \Free{} for using none or other AI tools than the free version of Gemini, and 9 participants were excluded in \Paid{} for using none or other AI tools than the paid version of Gemini.

The demographics of the remaining 159 participants can be found in Table~\ref{tab:demographics}.
Our sample involved 91 freelance and 37 industrial developers. 
147 participants were men, 9 were women, and 3 did not disclose their gender, reflecting typical gender distribution in software development as found in the SO Developer Survey~\cite{StackOve32:online} with only 5.17\% of respondents identified as women.
The ages ranged from 18 to 54 with a mean age of 28.
95 had a bachelor's degree, and 44 held a master's or equivalent. 
Participants came from 42 countries, including Pakistan (33), India (27), the USA (13), and Nigeria (11).

\begin{table}[htbp]
  \centering
  
  \caption{Demographics of the 159 Participants.}
  \vspace{-1em}
  \small
  \begin{tabularx}{0.5\textwidth}{X}
    \toprule
    \rowcolor{gray!10} \textbf{Gender} \\
    Man: 147, Woman: 9, Prefer not to disclose: 3 \\
    \rowcolor{gray!10} \textbf{Age*} \\
    min = 18.0, max = 54.0, mean = 28.26, median = 27.0, sd = 6.21\\
    \rowcolor{gray!10} \textbf{Educational Qualification}\\
    Bachelor Degree: 95, Masters Degree or equivalent Diploma: 44,
    School degree: 6, Professional training: 4, Doctoral degree (Dr./PhD): 3, No school degree: 3, Other: 4 \\
    \rowcolor{gray!10} \textbf{Main Occupation}\\ 
    Freelance developer: 91, Industrial developer: 37, Graduate student: 9,  Undergraduate student: 9, Academic researcher: 2, 
    Other: 11\\
    \rowcolor{gray!10} \textbf{Country of Residence}\\ 
    PK: 33, IN: 27, US: 13, NG: 11, GB: 9, KE: 7, EG: 6, CA: 4, ET: 4, UA: 4, NP: 3, MA: 3, MY: 2, AE: 2, ES: 2, DE: 2, DZ: 2, Other: 25\\
    \rowcolor{gray!10}\textbf{General Development Experience [years]*}\\
    min = 1.0, max = 40.0, mean = 6.53, median = 5.0, sd = 4.88\\
    \rowcolor{gray!10} \textbf{Python Experience [years]*}\\
    min = 1.0, max = 25.0, mean = 4.63, median = 4.0, sd = 3.17\\
    \rowcolor{gray!10} \textbf{Commonly Used AI Tools}\\ 
    ChatGPT Free Version: 101, Gemini Free Version: 57,
    ChatGPT Paid Version: 39, GitHub Copilot: 34, Gemini Paid Version: 8,
    Visual Studio IntelliCode: 8, Tabnine Free Version: 3, None: 1, Other: 9\\
    \rowcolor{gray!10} \textbf{Security Experience}\\
    No experience: 63, Security course/training: 45, 
    Developed security \\applications/implemented security measures: 42,
    Worked on IT security in spare time: 14, Worked at IT security-related companies: 10,
    Certificate in IT security: 4, Degree in IT security: 3, Other: 4 \\
    \bottomrule
    * = There were no significant differences between the groups.\\ 
  \end{tabularx}
  \label{tab:demographics}
\end{table}

\subsection{Evaluation}\label{sec:evaluation}

We analyzed whether the submissions fulfilled our functional and security requirements as follows. 

\textbf{Code Analysis:}
Two researchers independently reviewed all submissions to assess both functional correctness and potential security vulnerabilities. As part of the security evaluation, the reviewers actively examined whether issues such as XSS, CSRF, SQL injection, or improper input validation might arise. When their assessments differed, the researchers engaged in a structured discussion to analyze the discrepancy and reach a consensus on the final judgment.
We used a binary score for (non-)existent vulnerabilities (0 or 1).
Finally, we checked all submissions with 2 static scanners, Bandit~\cite{PyCQAban31:online} and SonarQube~\cite{CodeQual5:online}, which are listed by OWASP~\cite{SourceCo79:online}. This enabled us to verify our findings and cross-check that nothing had been overlooked. SonarQube is a widely used solution for analyzing code quality and detecting bugs and vulnerabilities, used by 7 million developers and over 400,000 organizations~\cite{CodeQual5:online}. Bandit is an open-source tool that analyzes Python code for security vulnerabilities. It is integrated into over 48,000 projects on GitHub~\cite{PyCQAban31:online}. 
Additionally, participants received one point for securely storing user passwords. For this, we expected participants to hash and salt passwords using state-of-the-art password hashing functions or schemes (e.g., bcrypt~\cite{provos_future-adaptable_1997}, scrypt~\cite{percival_stronger_2009}, Argon2~\cite{biryukov_argon2_2015}), according to~\cite{naiakshina_why_2017, hatzivasilis_password_2015}.

Two researchers manually evaluated the participants' submissions and resolved conflicts by discussion. 
The final security score ranged from 0 to 5.

\textbf{Open Survey Questions:} 
Our follow-up survey included open-ended questions about challenges faced during implementation, support and difficulties with Gemini, factors influencing trust and mistrust in AI, and the security best practices employed.
Two researchers conducted a thematic analysis~\cite{boyatzis_transforming_1998} of the responses to open-ended questions. First, they developed an initial codebook based on the responses from 5 random participants. Following this, each researcher independently coded the responses, regularly meeting to discuss and refine emerging themes and categories. During these discussions, they collaboratively added or merged codes, resulting in the final version of the codebook. After both researchers completed coding, they resolved discrepancies through discussion, reaching full agreement.
This approach aligns with the guidance of Braun and Clarke~\cite{Gotquest27:online, braun_starting_2022}.
The final codebook can be found in the Appendix~\ref{codebook}.

\subsection{Methodology: Statistical Analysis and Hypothesis Testing}
\label{sec:analysis}

We derived hypotheses from our research questions and used standard statistical hypothesis tests to examine the following 4 main hypotheses in our study:

\begin{itemize}
    \item H1: There is a significant effect of the programming experience on the security score.
    \item H2: There is a significant effect of security experience on the security score.
    \item H3: There is a significant difference in the security scores among \Free{},  \Paid{}, and \No{}.
    \item H4: There is a significant effect of the tool (\Free{}, \Paid{}) on the trust in the tool.
\end{itemize}

To address the first three hypotheses, we modelled the binary \textit{Security Score} (the number of passed security test cases out of $K\!=\!5$) with a logistic regression (binomial GLM) that simultaneously includes all hypothesised predictors.  The full model is  

\begin{flalign*}
\text{logit}\bigl(\Pr(\text{Success})\bigr)
  &= \beta_{0}
   + \beta_{1}\,\text{Group}_{\text{Free-AI}}
   + \beta_{2}\,\text{Group}_{\text{Paid-AI}} &\\
&\quad
   + \beta_{3}\,\text{Security Experience} &\\
&\quad
   + \beta_{4}\,\text{Programming Experience} &\\
&\quad
+ \beta_{5}\,\text{Security Self-Efficacy} &\\
&\quad
+ \beta_{6}\,\text{Workload} &\\
&\quad
   + \varepsilon &%
\end{flalign*}

where  

\begin{itemize}
  \item \texttt{Group} is a three‑level factor (No‑AI, Free‑AI, Paid‑AI) coded with the control (No‑AI) as reference;  
  \item \texttt{Security Experience} is a binary indicator of self‑reported prior security experience (0 = no experience, 1 = experience);  
  \item \texttt{Programming Experience}, \texttt{Security Self-Efficacy} and \texttt{Workload} are continuous, z‑standardised measures.  
\end{itemize}

All predictors are entered simultaneously, thus each coefficient reflects the unique contribution of the corresponding variable while holding the others constant.  This single‑model approach directly yields effect‑size estimates (log‑odds $\beta$, exponentiated to odds‑ratios, OR) that answer the hypotheses without the need for a series of separate tests.

\paragraph{Model diagnostics.}  Prior to inference, we examined over‑dispersion by comparing the residual deviance to its degrees of freedom; the ratio was well below the critical value of 1.5, indicating that the binomial assumption holds.  When over‑dispersion had been detected, we would have refitted the model with a quasibinomial family, but this was not required.

\paragraph{Hypothesis testing.}  For each categorical predictor (\texttt{Group}) we performed a Type‑II likelihood‑ratio test ($\chi$²) to evaluate the null hypothesis $H_{0}:\beta_{j}=0$ (no effect) against the alternative $H_{1}:\beta_{j}\neq0$.  Continuous predictors (\texttt{Programming Experience, Security Self-Efficacy, Workload}) are tested with Wald $z$‑statistics. All tests used a two‑tailed $\alpha$‑level of .05; marginally significant effects ($p$ between .05 and .10) are reported as trends.

\paragraph{Effect‑size estimation.}  Model coefficients were exponentiated to obtain odds‑ratios with 95\% confidence intervals (via the Wald method).  To aid interpretability we derived marginal means (estimated probabilities) for each level of \texttt{Group} using the `emmeans` package, back‑transforming from the logit to the probability scale.  Pairwise differences between AI conditions were examined with Tukey‑adjusted odds‑ratio contrasts to control the family‑wise error rate.

\paragraph{Model fit.}  Goodness‑of‑fit was summarised with pseudo‑$R^{2}$ indices (Nagelkerke’s $R^{2}$=0.11) and the Hosmer–Lemeshow test\\ ($\chi$²\textsubscript{8}=5.3, $p$=0.73) to confirm adequate calibration. 
In short, the logistic regression framework allows us to test the main effects of programming experience, security experience, and Gemini assistance (RQ1–RQ3) in a single, statistically rigorous model, while the Type‑II LR tests, odds‑ratio reporting, and post‑hoc Tukey contrasts provide the inferential basis for accepting or rejecting the corresponding hypotheses.

For the fourth research question, we calculated descriptive statistics (means and standard deviations) for each condition. To test the hypothesis that trust levels differ between the \Free{} and \Paid{} groups, we compared the two independent samples using the non‑parametric Mann‑Whitney U test (appropriate for ordinal data and non‑normal distributions). The test was performed separately for each trust item, yielding U statistics and two‑tailed p‑values.

\subsection{Ethics}
Our university did not have an institutional review board (IRB) at the time of this study.
However, we complied with the General Data Protection Regulation (GDPR) cleared with our data protection officer. To protect participants’ privacy, we minimized the collection of personal data and only collected information necessary for the study. Participants were informed about the study's procedure and provided with a consent form, including our data policies and the study information, which also provided contact details for the research team and the data protection officer. Participation was voluntary, and participants were asked to agree to the consent form before proceeding. Participants were allowed to ask questions at any time during the study and withdraw from the study at any time without facing any consequences.
At the end of the study, we debriefed and informed all participants about the purpose of the study.

\subsection{Limitations}

As with most user studies, this study has several limitations that should be considered when interpreting its findings.
First, we recruited freelance developers for this study, which may limit the generalizability of our findings to other populations. Developers employed by companies or recruited through other platforms may behave differently. 
Second, the programming task in our study was intentionally designed to be susceptible to various vulnerabilities in Python. Different programming tasks or the use of other programming languages may have led to different outcomes.
Third, we focused exclusively on the free and paid versions of the AI tool Gemini. 
Further research is needed to explore how results vary across different tools. 
Fourth, following best practices that recommend instructing participants not to use LLMs when they are not permitted~\cite{prolific_llm_detection}, we instructed participants either not to use any AI tools or to use only Gemini, depending on their condition. While we could not verify compliance, all participants were informed that they would be compensated, so we have no indication of incentives to misreport. 
Fifth, due to the remote study design, our analysis was limited to participants’ code submissions. 
Thus, we could not verify how prompting behavior influenced the security score.
Finally, the observed effect size is modest, and the model explains only a limited portion of the variance in the security score. The sample size may have limited our ability to detect smaller effects, suggesting that additional factors could have influenced security scores.

\begin{table*}[t]
  \centering
  \caption{Overview of Task Statistics.}
  \setlength{\tabcolsep}{2.8pt}
  \small
  \begin{tabular}{lccc}
    \toprule
    \textbf{Time in Minutes} & mean = 143.89, median = 120.0 & min = 1.0, max = 800.0 & sd = 128.57\\
    \rowcolor{gray!10}\textbf{Resources Used} & \multicolumn{3}{l}{\begin{tabular}[t]{@{}l@{}}Official documentation: 90, StackOverflow: 84, GitHub: 44, Tutorial Websites: 37,\\ YouTube: 18, W3Schools: 10, Codecademy: 1, Other: 27 \\\end{tabular}} \\
    \textbf{SUS Gemini Free} & mean = 76.1, median = 77.5 & min = 32.5, max = 100 & sd = 14.84\\
    \rowcolor{gray!10}\textbf{SUS Gemini Paid} & mean = 77.1 median = 77.5 & min = 50.0, max = 100  & sd = 14.15\\
    \textbf{Helpfulness Gemini Free*} & mean = 4.68, median = 5.0 & min = 2.0, max = 7.0 & sd = 1.33\\
    \rowcolor{gray!10}\textbf{Helpfulness Gemini Paid*} & mean = 4.92, median = 5.0 & min = 2.0, max = 7.0 & sd = 1.44\\
    \textbf{Number of Prompts Gemini Free} & mean = 11.62, median = 5.0 & min = 1.0, max = 80.0 &  sd = 16.51\\
    \rowcolor{gray!10}\textbf{Number of Prompts Gemini Paid} & mean = 29.64, median = 10.0 & min = 1.0, max = 500.0 & sd = 74.08\\
    \textbf{\% of Adopted Suggestions Gemini Free} & mean = 57.17, median = 60.0 & min = 0.0, max = 100.0 & sd = 22.73\\
    \rowcolor{gray!10}\textbf{\% of Adopted Suggestions Gemini Paid} & mean = 56.42, median = 60.0 & min = 10.0, max = 100.0 & sd = 25.88\\
    \bottomrule
    \multicolumn{4}{c}{* = Likert Scale, 1: Not Helpful - 7: Very Helpful.} \\ %
  \end{tabular}
  \label{tab:task}
\end{table*}

\begin{table*}[htbp]
  \centering
  \caption{Estimated Odds Ratios and 95\% CIs from the Binomial GLM}
  \begin{tabular}{lllllll                 }
    \toprule
    \textbf{Term} & \textbf{Estimate} & \textbf{Std.error} & \textbf{Statistic} &
    \textbf{p.value} & \textbf{Conf.low} & \textbf{Conf.high} \\
    \midrule
    (Intercept)       & 0.64 & 0.173 & -2.61 & 0.0090 & 0.45 & 0.89 \\
    \rowcolor{gray!10}\Free      & 1.31 & 0.199 &  1.34 & 0.180  & 0.88 & 1.93 \\
    \Paid      & 1.54 & 0.192 &  2.24 & 0.0252 & 1.06 & 2.25 \\
    \rowcolor{gray!10}Programming Experience          & 1.23 & 0.0818 & 2.55 & 0.0107 & 1.05 & 1.45 \\
    Security Self-Efficacy         & 0.92 & 0.0869 & -0.94 & 0.346  & 0.78 & 1.09 \\
    \rowcolor{gray!10}Workload       & 1.08 & 0.0804 &  0.96 & 0.337  & 0.92 & 1.27 \\
    Security Experience   & 0.77 & 0.178 & -1.44 & 0.151  & 0.55 & 1.10 \\
    \bottomrule
  \end{tabular}
  \label{tab:glm-odds}
\end{table*}

\section{Results}
\label{sec:results}

This section describes the results of the 4 research questions. Following the evaluation of our quantitative results, we present an analysis of our qualitative findings for each research question to provide further insight and explanation.
We report individual statements and results by referencing the \No{} participants with \emph{N}, the \Free{} participants with \emph{F}, and the \Paid{} participants with \emph{P}.
Participants completed the task with a mean time of 144 minutes. There was no statistical difference in the completion time between the three groups. 
An overview of the general descriptive statistics is presented in Table~\ref{tab:task}. %

\textbf{Functionality:} Of the 159 participants (\No{} n = 53, \Free{} n = 53, \Paid{} n = 53), 138 submitted functional solutions (\No{} n = 47, \Free{} n = 41, \Paid{} n = 50). 

In the following, we report our security analysis of participants' code by considering only functional solutions.

We modeled participants’ security score as a binomial outcome using logistic regression, with the number of correct items out of K = 5 per participant as the dependent variable. The model included Group (three levels) and four covariates entered as main effects: standardized overall programming experience, standardized self-efficacy, standardized workload, and prior security experience. We report likelihood-ratio tests for each term and odds ratios with 95\% CIs; adjusted marginal means by group are presented on the 0–5 score scale (i.e., probability × 5). Given the high correlation between Python-specific and overall programming experience (r = 0.78), Python experience was omitted from the final model to prevent multicollinearity; overall programming experience was retained. Table~\ref{tab:glm-odds} provides an overview of the estimated odds ratios and their 95\% confidence intervals, showing how each variable is associated with changes in the odds of the outcome.

\begin{table}[t]
  \centering
  \caption{Pseudo‑$R^2$ Statistics for the Final Model.}
  \label{tab:pseudoR2}
  \begin{tabular}{lccc}
    \toprule
    McFadden & Cox \& Snell & Nagelkerke \\
    \midrule
    0.035 & 0.10 & 0.103 \\
    \bottomrule
  \end{tabular}
\end{table}

\textbf{Model Diagnostics:}
Over‑dispersion was assessed via the Pearson $\chi^2$ statistic.  The dispersion factor $\hat{\phi}=0.87$ indicated \emph{no} serious over‑dispersion, so a standard binomial GLM was retained (the beta‑binomial alternative was not triggered). Pseudo‑$R^2$ values (Table~\ref{tab:pseudoR2}) show modest explanatory power (McFadden $R^2\!=\!0.035$).

\subsection{RQ1 - Programming Experience and Software Security}
\label{sec:rq1}

\paragraph{Quantitative Analysis.}
\textbf{Regression Analysis:}
A one-standard-deviation increase in overall programming experience raised the odds of a higher security score by a factor of $\mathbf{1.43}$ (95 \% CI $[1.14,\,1.81]$, $p\!=\!0.002$).  In probability terms, the predicted security success rate increased from $30\%$ (mean experience) to $38\%$ at $+1$ SD. A one‑SD increase in programming experience raised the odds of a higher security score by 23\%.

\begin{figure}[t]
\centering 
\includegraphics[width=0.5\textwidth]{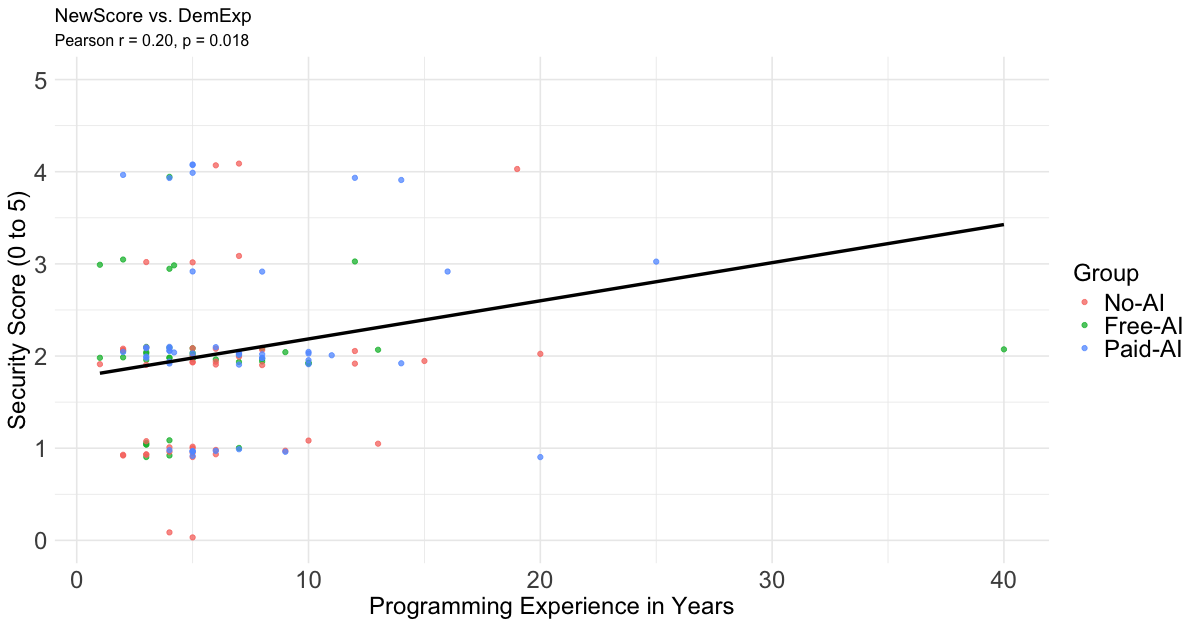}
\caption{Correlation Programming Experience and Security Score.}
\label{fig:rq3-programmingexperience}
\end{figure}

\textbf{Descriptive Values:}

Figure~\ref{fig:rq3-programmingexperience} visualizes the association between the security score and programming experience: points show individual participants (colored by group), with a least-squares trend line. The relationship was approximately linear (Pearson r = 0.2, p = 0.018). Table~\ref{tab:summary-stats-programming-exp} reports descriptive statistics for Security Score and Programming Experience by group (mean, SD, n, min, and max).

\begin{table}[t]
\centering
\caption{Descriptive Statistics for Programming Experience (Years) and Security Scores by AI Group.}
\small
\resizebox{0.35\textwidth}{!}{
\begin{tabular}{lcccccc} 
\toprule 
Group & Count & Mean & SD & Min & Max \\
\midrule 
\multicolumn{6}{c}{\textbf{General Programming Experience}} \\
\midrule 
\No{} & 47 & 6.70 & 4.68 & 1 & 22 \\ 
\rowcolor{gray!10}\Free{} & 41 & 6.25 & 6.23 & 1 & 40 \\
\Paid{} & 50 & 7.16 & 4.52 & 2 & 25 \\ 
\midrule 
\multicolumn{6}{c}{\textbf{Security Score}} \\
\midrule 
\No{} & 47 & 1.81 & 1.17 & 0 & 5 \\
\rowcolor{gray!10}\Free{} & 41 & 2.07 & 0.79 & 1 & 5 \\
\Paid{} & 50 & 2.26
 & 1.10
 & 0 & 5 \\
\bottomrule 
\end{tabular}
}
\label{tab:summary-stats-programming-exp}
\end{table}

\paragraph{Qualitative Analysis.}
\textbf{Validation by Experienced Developers:}
Developers reported using their knowledge and experience to evaluate, verify, and correct Gemini's output. Rather than blindly trusting the AI, they reported using it as a starting point, cross-checking its outputs with their experience and other resources. 
For instance, P16 noted: \contextquote[P16]{P16}{I have background knowledge of backend development with Flask and had a general idea of where the LLM was going in development}. F56 shared: \contextquote[F56]{F56}{with my experience and skill set, I can strategically test, verify, and take over to ensure the final code for a given task meets the needs of the project}.
This sentiment was echoed by 6 participants in \Free{} (F05, F06, F34, F36, F56, F67) and 5 participants in \Paid{} (P16, P26, P32, P56, P59), who described their ability to assess Gemini’s suggestions due to their knowledge or experience. In summary, the fact that experienced developers verified Gemini’s output might help explain the association between programming experience and higher security scores, suggesting that programming experience remains important for critically evaluating and integrating Gemini’s suggestions.

\begin{summary}{RQ 1 -- Summary}
We found that general programming experience %
significantly improved code security. Developers with more years of experience wrote more secure code. Many participants emphasized the importance of evaluating AI-generated code using their own experience.
\end{summary}

\subsection{RQ2 - Security Experience and Software Security}
\label{sec:rq2}
\paragraph{Quantitative Analysis.}
We divided the participants into 2 groups: those who reported having previous security experience and those without.
Security experience included security education, certifications, implementation of security measures, work in security-related companies, or personal security projects.

Developers who reported prior security experience tended to have lower odds of a higher security score ($\mathbf{0.75}$, 95 \% CI $[0.55,\,1.01]$, $p\!=\!0.058$).  Although the effect narrowly missed conventional significance ($\alpha=0.05$), the direction is noteworthy: participants with security experience produced software that, on average, had lower security scores (predicted probability $=0.27$) than those without such experience (predicted probability $=0.30$). 

The mean security score for participants without prior security experience was 2.20, compared to 1.95 for those with experience. Table~\ref{tab:summary-stats-security-exp-ai} further shows that participants without security experience achieved higher mean security scores when using Gemini than participants with experience. When no AI tool was used, participants with and without prior security experience showed nearly identical mean scores (1.80 vs. 1.81). In contrast, when Gemini assistance was available, participants without prior security experience achieved higher mean scores than those with security experience, indicating that Gemini may have mitigated differences related to security experience within the scope of this analysis. Table~\ref{table:summary_stats_selfefficacy} presents the descriptive statistics for the participants’ SSD-SES scores, which were comparable across groups.

\begin{table}[t]
\centering
\caption{Descriptive Statistics of Security Scores by Security Experience by AI Group.}
\small
\resizebox{0.5\textwidth}{!}{
\begin{tabular}{ccccccc}
\toprule
Security Experience & Group & Mean & Variance & Min & Max & Count \\
\midrule
No Experience & \No{} & 1.80 & 1.22 & 1 & 5 & 20 \\
\rowcolor{gray!10}No Experience & \Free{} & 2.12 & 0.49 & 1 & 4 & 17 \\
No Experience & \Paid{} & 2.68 & 1.67 & 0 & 5 & 19 \\
\rowcolor{gray!10}No Experience & All & 2.20 & 1.25 & 0 & 5 & 56 \\
Experience & \No{} & 1.81 & 1.54 & 0 & 5 & 27 \\
\rowcolor{gray!10}Experience & \Free{} & 2.04 & 0.74 & 1 & 5 & 24 \\
Experience & \Paid{} & 2.00 & 0.80 & 1 & 4 & 31\\
\rowcolor{gray!10}Experience & All & 1.95 & 1.01 & 0 & 5 & 82 \\
\bottomrule
\end{tabular}
}
\label{tab:summary-stats-security-exp-ai}
\end{table}

\begin{table*}[t]
\centering 
\caption{Summary Statistics for Security Self-efficacy by Group}
\begin{tabular}{lccccccccccc} 
\toprule 
& & & & & & & \multicolumn{5}{c}{Participants per Score Range} \\
Group & Count & Mean & Variance & Min & Max & Median & 0-1 & $>$1-2 & $>$2-3 & $>$3-4 & $>$4-5 \\ 
\midrule 
\No{} & 47 & 3.31 & 0.90 & 1.33 & 4.80 & 3.60 & 0 & 6 & 12 & 16 & 13 \\ 
\rowcolor{gray!10}\Free{} & 41 & 3.44 & 0.81 & 1.20 & 4.73 & 3.67 & 0 & 5 & 6 & 18 & 12 \\ 
\Paid{} & 50 & 3.57 & 0.58 & 2.00 & 4.93 & 3.73 & 0 & 1 & 13 & 20 & 16 \\
\bottomrule 
\end{tabular}
\label{table:summary_stats_selfefficacy}
\end{table*}

\paragraph{Qualitative Analysis.}
\textbf{Leveraging Security Experience:}
F44 emphasized the critical role of experience when reviewing AI-generated code: \contextquote[F44]{F44}{auditing for security against today's high-end skilled cyber attacker is necessary with assistance of experienced developer}.
Similarly, P27 and P32 stress the value of their personal experience in identifying vulnerabilities: \contextquote[P27]{P27}{Based on personal experience, I can identify weak points}, \contextquote[P32]{P32}{As I have experience so I can know easily if the presented code makes sense or not. Or if there are any security issues}.
P16 highlights the importance of experience, particularly in security:
\contextquote[P16]{P16}{I also have background knowledge of password storage [...]. At each step, I tested the generated code by Gemini to confirm that it met requirements in terms of functionality and security. It is not safe to trust code generated by Gemini blindly}.
In contrast, F08 described a different perspective, trusting the code generated by Gemini precisely because they lacked sufficient security knowledge: \contextquote[F08]{F08}{I'm not knowledgeable enough about security implementations to know if the code provided is insecure}. This might indicate that developers without prior security experience may
have relied more heavily on Gemini or formulated more targeted
security-related prompts, which may help contextualize the quantitative finding that participants without prior security experience achieved slightly higher security scores when using Gemini.

\begin{summary}{RQ 2 -- Summary}
We observed no significant effect of security experience on the security score. However, developers with no prior security experience achieved slightly higher security scores when using Gemini compared to participants without experience who did not use AI. Qualitative responses suggest that some developers with security experience critically evaluated Gemini’s suggestions, whereas some less experienced participants relied more directly on Gemini's output. %
\end{summary}

\subsection{RQ3 - \Paid{} vs. \Free{}}
\label{sec:rq3}

\paragraph{Quantitative Analysis.}
\textbf{Main Effect of AI Group:}

The omnibus Type‑II LR test for \texttt{Group} was marginally significant  ($\chi^2_{(2)}=5.12$, $p=0.077$), suggesting a trend toward differential impact of Gemini assistance on security scores. Post‑hoc pairwise contrasts (Table~\ref{tab:groupContrasts}) confirm this pattern.

\begin{table}[htbp]
  \centering
  \caption{Pairwise Odds‑Ratio Comparisons b
  Between AI Groups (Adjusted With Tukey’s Method).}
  \begin{tabular}{lcc}
    \toprule
    Contrast & Odds Ratio & $p$ \\
    \midrule
    \No{} vs. \Free{} & 0.766 & 0.3732 \\
    \rowcolor{gray!10}\No{}{} vs. \Paid{} & 0.650 & 0.0650 \\
    \Free{} vs. \Paid{} & 0.849 & 0.6772 \\
    \bottomrule
  \end{tabular}
  \label{tab:groupContrasts}
\end{table}

\textbf{Estimated Probabilities:}

Marginal means back‑transformed to the probability scale (Table~\ref{tab:marginalProbs}) show a monotonic increase from the control condition (\No{}) to \Paid{}, but none of the differences reach statistical significance. After adjusting for experience, self‑efficacy, workload, and prior security background, participants who used the paid version achieved a ~10\% absolute gain in security scored relative to the \No{} Group (95\% CI: 5.8\%–14.9\%; p = 0.06). Table~\ref{tab:summary-stats-nasa-tlx} displays the summary statistics for the NASA-TLX. Notably, the \Paid{} group had lower workload scores compared to the other 2 groups.

\begin{table}[htbp]
  \centering
  \caption{Estimated Probability of a Successful Security Outcome by AI Group (Averaged Over Security Experience).}
  \begin{tabular}{lcc}
    \toprule
    Group & Probability & 95\,\% CI \\
    \midrule
    \No{}   & 0.359 & [0.299, 0.424] \\
    \rowcolor{gray!10}\Free{} & 0.422 & [0.355, 0.492] \\
    \Paid{} & 0.462 & [0.400, 0.526] \\
    \bottomrule
  \end{tabular}
  \label{tab:marginalProbs}
\end{table}

\begin{table*}[h]
\centering
\caption{Summary Statistics for the NASA-TLX by Group}
\resizebox{0.9\textwidth}{!}{
\begin{tabular}{cccccccccccccccc}
\toprule
& & & & & & & \multicolumn{9}{c}{Participants per Score Range} \\
Group & Count & Mean & Variance & Min & Max & Median & 0-10 & 11-20 & 21-30  & 31-40 & 41-50 & 51-60 & 61-70 & 71-80 & 81-90\\
\midrule
\No{}    & 47 & 41.49 & 341.12 & 6 & 84.00 & 40.67 & 2 & 5 & 6 & 10  & 9 & 11 & 1 & 2 & 1\\
\rowcolor{gray!10}\Free{}    & 41 & 40.40 & 376.67 & 2 & 86.67 & 42.67 & 2 & 5 & 6 & 6  & 10 & 5 & 4 & 2 & 1\\
\Paid{} & 50 & 33.44 & 448.50 & 7 & 90.00 & 28.00 & 4 & 11 & 14 & 6  & 5 & 2 & 5 & 1 & 2\\
\bottomrule
\end{tabular}
}

\label{tab:summary-stats-nasa-tlx}
\end{table*}

\begin{figure}[t]
\centering 
\includegraphics[width=0.5\textwidth]{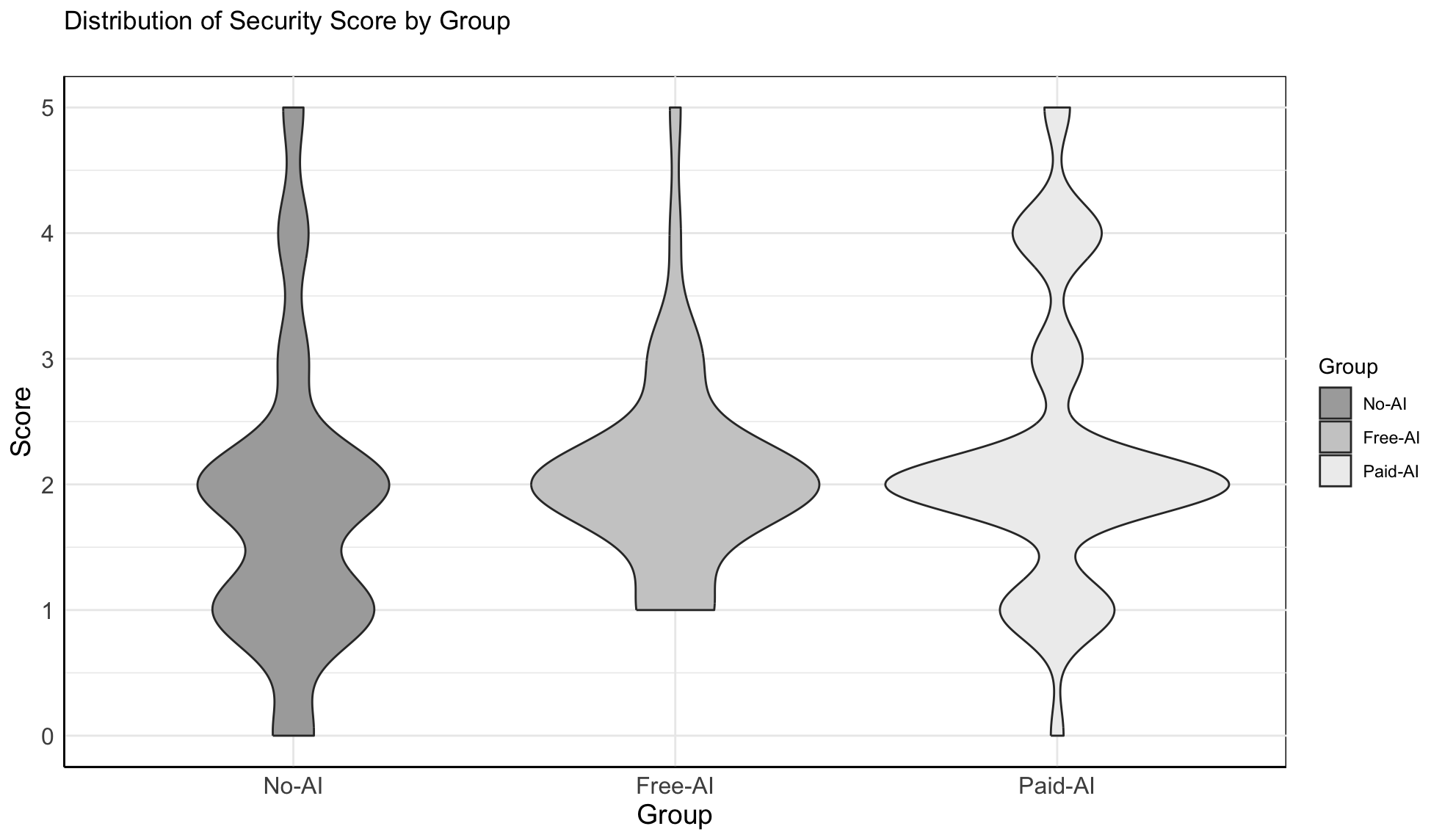}
\caption{Violinplot for Security Score Divided by Group (0 = lowest, 5 = highest security score).}
\label{fig:rq1-violinplot-totalsecurityscore}
\end{figure}

\textbf{Security:}
We evaluated code security on a scale from 0-5 (see Section~\ref{sec:evaluation}).
The mean values of the security score for each group were relatively close, with \Free{} (2.07) and \Paid{} (2.26) being similar, while \No{} is slightly lower (1.81). %
Table~\ref{tab:score-percentage} shows the percentages of the individual vulnerabilities in the security score. Notably, password-storage security improved by 21.7\% with Gemini assistance, and SQL injection vulnerabilities improved by 10.5\%. In contrast, XSS vulnerabilities showed only a 3.5\% improvement, while CSRF and improper input validation exhibited almost no change with Gemini assistance.
The distributions of the scores for each group are visualized in Figure~\ref{fig:rq1-violinplot-totalsecurityscore}, using violin plots to illustrate the spread and density of the data. \No{} shows the highest distribution around the security score of 1 with 16 participants and around 2 with 19 participants. \Free{} shows a high number of 27 participants with a score of 2 and 5 participants with a score of 3. \Paid{} has the highest number of participants with also 27 at around 2 points, but also 9 participants with 1 and 7 participants with 4.

\begin{table*}[htbp]
\centering
\caption{Percentage of Vulnerabilities by Group.}
\scriptsize The ``Average AI'' column shows the average percentage of vulnerabilities for participants who used AI (\Free{} and \Paid{}), followed by the difference to \No{}.
\small
\begin{tabular}{lccccc}
\toprule
\textbf{Category} & \textbf{\No{}} & \textbf{\Free{}} & \textbf{\Paid{}} & \textbf{Average AI} & \textbf{Difference \No{} to AI}\\ 
\midrule
\rowcolor{gray!10}Insecure Password Storage & 40.4 & 17.1 & 20 & 18.68 & -21.7 \\
SQL Injection (CWE-89) Vulnerable & 21 & 7 & 14 & 10.5 & -10.5 \\
\rowcolor{gray!10}XSS (CWE-79) Vulnerable & 79 & 85 & 66 & 75.5 & -3.5 \\
CSRF (CWE-352) Vulnerable & 96 & 93 & 96 & 94.5 & -1.5 \\
\rowcolor{gray!10} Improper Input Validation
(CWE-20) Vulnerable & 83 & 90 & 78 & 84 & 1 \\
\bottomrule
\end{tabular}
\label{tab:score-percentage}
\end{table*}

\paragraph{Qualitative Analysis.}
\textbf{Security Assistance From Gemini:}
Five participants in \Paid{} (P31, P44, P46, P51, P62) and 3 participants in \Free{} (F05, F35, F54) mentioned using Gemini to get general suggestions about security (e.g., \contextquote[F35]{F35}{it suggested some techniques to improve the security of the website}).
Four participants from \Free{} (F05, F35, F54, F65) and 3 from \Paid{} (P01, P27, P62), reported using Gemini to identify vulnerabilities.
For example, P01 explained their process: \contextquote[P01]{P01}{First I query Gemini AI to complete the tasks provided with task and supporting code. Then wrote another query to rewrite the task (functions) to optimize it and make it secure}. 
P62 shared that Gemini \contextquote[P62]{P62}{helped talk through potential security issues}, emphasizing its role in identifying vulnerabilities.

\textbf{Security Challenges With Gemini:} Eleven participants in \Paid{} and 6 in \Free{} highlighted various ways in which Gemini hindered them in writing secure code. Two participants (P27, P43) noted that the security suggestions were not optimal 
(e.g., \contextquote[P27]{P27}{Sometimes it correctly points on a possible security issue, but solution is not accurate or being done in a more convenient way than it suggests}).
The quality of explanations for security suggestions was criticized by 5 participants (F35, F64, F67, P56, P57). P57 felt that \contextquote[P57]{P57}{The explanation wasn't convincing and/or the risk of trusting any wrong code was very high in cases such as password hashing or verification}.
 
Three participants (F59, P43, P58) noted that when asking Gemini for security suggestions, \contextquote[P43]{P43}{it often provides code with errors}.
The generated code not adhering to best security practices was mentioned by F64 and P01, while F44 and P49 noted that Gemini was more likely to introduce new vulnerabilities in the generated code.

\textbf{Credential Storage:}
Four participants in \Paid{} (P30, P35, P46, P57) and 4 in \Free{} (F02, F23, F39, F64) reported using Gemini for assistance with hashing. For instance, F02 mentioned that Gemini \contextquote[F02]{F02}{Just helped [them] decide over hashing methods}, highlighting its assistance in making informed decisions. P57 appreciated how Gemini introduced approaches they might not have considered on their own: \contextquote[P57]{P57}{It has provided me approaches that I probably wouldn't have thought of beforehand. Its implementation of hash for making the application more secure was helpful}. Three participants (F31, P07, P16) mentioned Gemini’s failure to incorporate security into its suggestions. P16 described how Gemini’s responses lacked crucial security measures, stating: \contextquote[P16]{P16}{The initial responses from Gemini on registration/login showed no hashing or salting of passwords and tried storing them in plain text}.
Outdated security suggestions were a concern for 3 participants (F64, P46, P48) (e.g., \contextquote[P46]{P46}{Gemini generated code based on older versions of bcrypt}).

\textbf{SQL-Injection:}
Two participants (F05, P36) reported using Gemini for SQL injection prevention.
P36 explained how Gemini corrected their misunderstanding of SQL parameter interpolation, preventing an SQL injection vulnerability in their code: \contextquote[P36]{P36}{I assumed that SQL params interpolation is done using '\%s' syntax, but Gemini suggested a syntax with '?'. I corrected it, but turns out I was wrong and Gemini was correct}. 
Notably, 19 participants indicated that they actively implemented measures to prevent SQL injections as a best practice. 3 participants (P36, P43, N01) reported general challenges with preventing SQL injections 
(e.g., \contextquote[N01]{N01}{parameterizing queries to prevent SQL injection and properly handling database connections}).
The relatively high number of participants mentioning preventing SQL injections, coupled with the frequency of actually mitigated vulnerabilities, suggests a greater awareness of SQL injection vulnerabilities than improper input validation, CSRF and XSS.

\textbf{Improper Input Validation:}
Nine participants mentioned mitigating improper input validation.
F35 felt that the security suggestions for 
improper input validation and hashing were confusing: \contextquote[F35]{F35}{One time Gemini suggested Form validation and hash password as suggestion I think that was bit confusing and I was not able to implement it properly}.
P56 mentioned their challenge with 
improper input validation: 
\contextquote[P56]{P56}{I used package validators for validating URLs, but the sample code didn't mention what kind of validation it really does and I had several failed tests before looking up more elaborate docs}.
Two further participants (N18, N52) reported challenges with improper input validation. N52 described challenges with validating URLs: \contextquote[N52]{N52}{if you inserted ``facebook.com'' or ``www.facebook.com,'' it would raise an error 'Invalid URL'. That's because the missing HTTPS, so it has to be something like this https://www.facebook.com […] This took a lot of my time and was frustrating}.

\textbf{CSRF:}
Only 4 participants (N07, F15, F29, P62) reported taking measures to prevent CSRF. Interestingly, no participants reported having challenges with CSRF prevention. Given that CSRF was also the most prevalent vulnerability, this suggests there is a low awareness of CSRF.

\textbf{XSS:}
Just 2 participants (N26, P62) mentioned addressing XSS. 
P62 mentioned preventing \contextquote[P43]{P43}{XSS, [...] as suggested by Gemini }.
Only one participant, P43 reported challenges with XSS: \contextquote[P43]{P43}{The challenge was during the security part when I wanted to avoid [...] XSS attacks}.

\begin{summary}{RQ 3 -- Summary}
No statistically significant differences were observed among developers using no AI, the free version, or the paid version of Gemini. %
Participants found Gemini helpful for security tasks, including general security advice, guidance on hashing, and identifying vulnerabilities. 
However, they noted challenges such as suboptimal or outdated suggestions, missed measures, unclear explanations, errors, and concerns about Gemini's suggestions introducing vulnerabilities.
\end{summary}

\subsection{RQ4 - Impact of Gemini Use on User Trust}

\begin{figure}[t]
\centering 
\includegraphics[width=0.5\textwidth]{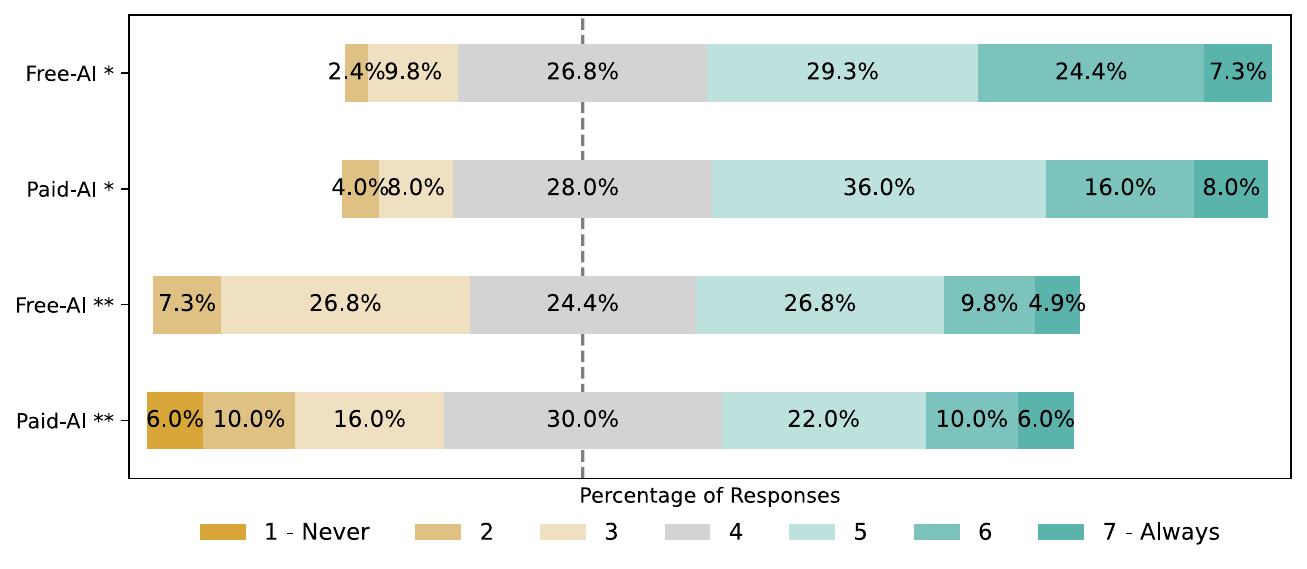}
\caption{Responses for *``I trust Gemini in general'' and **``I trust Gemini to generate secure code''.}
\label{fig:trust-combined}
\end{figure}

\begin{figure}[t]
\centering 
\includegraphics[width=0.5\textwidth]{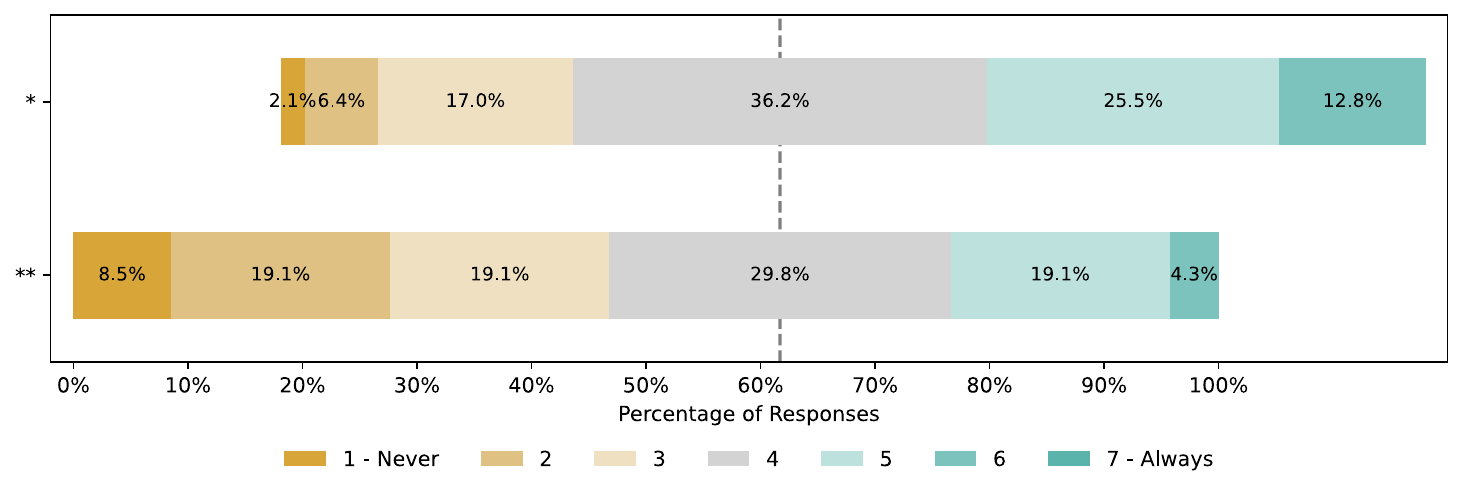}
\caption{Responses for *``Would you trust AI in general?'' and **``Would you trust AI to generate secure code?'', Asked for \No{} Group}
\label{fig:trust-hypothetical}
\end{figure}

\paragraph{Quantitative Analysis.}
\textbf{Trust in AI:}
We asked participants if they ``trust Gemini in general'' and to ``generate secure code'' on a Likert Scale from 1: Never - 7: Always (see Figure~\ref{fig:trust-combined}). Participants in the \No{} group were asked if they would trust AI (see Figure~\ref{fig:trust-hypothetical}). The mean scores for general trust in Gemini in both groups were similar, with \Free{} having a slightly higher mean (4.85) than \Paid{} (4.76).
The results for trust in Gemini to create \emph{secure} code were also similar. However, the mean values were slightly lower than for the general trust, 4.20 for \Free{} and 4.06 for \Paid{}.

\begin{figure}[t]
\centering 
\includegraphics[width=0.5\textwidth]{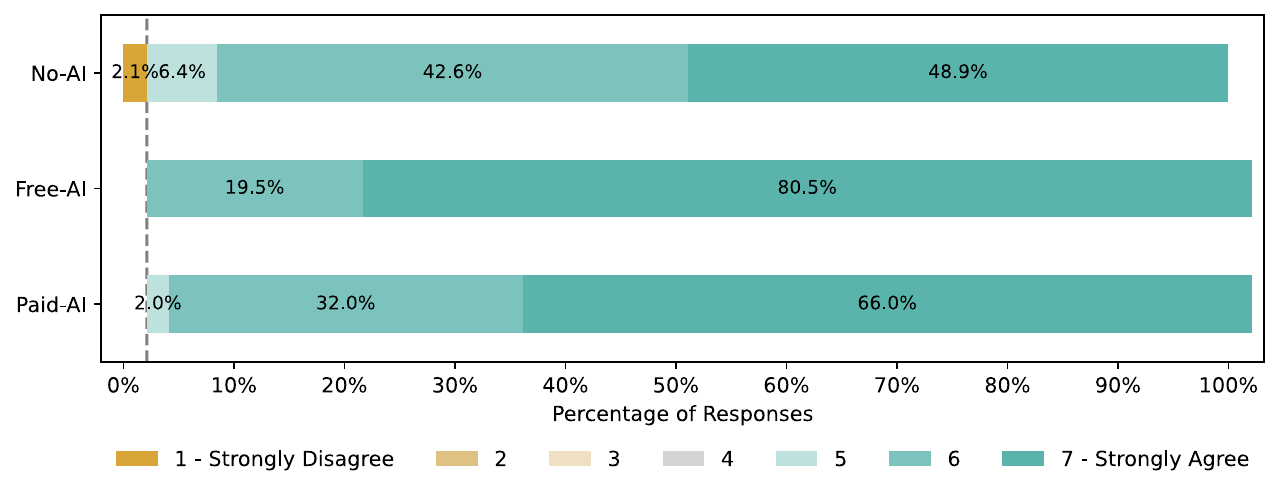}
\caption{Responses for ``I believe I have solved this task functionally correctly.''}
\label{fig:functionally}
\end{figure}

\begin{figure}[htbp]
\centering 
\includegraphics[width=0.5\textwidth]{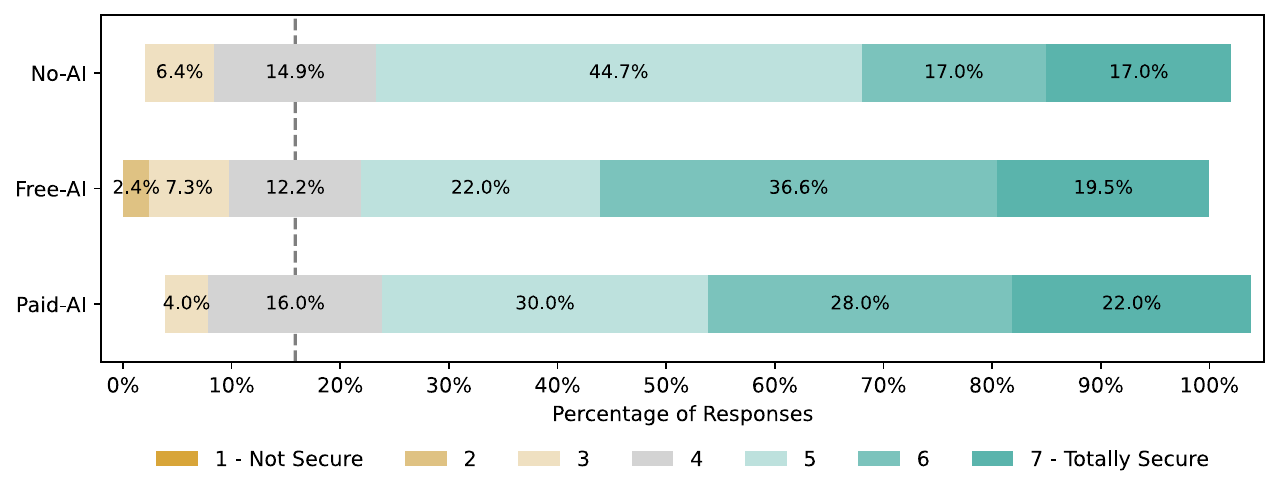}
\caption{Responses for ``How would you rate the level of security of your solution for the task?''}
\label{fig:secure}
\end{figure}

\textbf{Trust in Solutions:} 
Similarly, we asked participants to rate their self-assessed functional correctness in Figure~\ref{fig:functionally} and self-assessed security in Figure~\ref{fig:secure}. The mean scores for the self-assessed functional correctness were 6.32 in \No{}, 6.80 in \Free{} and 6.64 in \Paid{}. The mean scores for the self-assessed security were 5.23 in \No{}, 5.41 in \Free{} and 5.48 in \Paid{}. The results indicate that participants trust the functionality of their code more than the security.
Table~\ref{table:ai-preference} summarizes the AI-preference responses for \No{}, indicating that the majority of participants in the \No{} group would have preferred to use AI.

\begin{table}[t]
\centering
\caption{Responses to AI Assistance Preference, Asked for \No{} Group}
\begin{tabular}{lc}
\toprule
AI Preference & Number of Participants  \\
\midrule
No         & 13 \\
\rowcolor{gray!10}Yes        & 34 \\
\bottomrule
\multicolumn{2}{c}{``Would you have appreciated assistance from an AI?''}
\end{tabular}
\label{table:ai-preference}
\end{table}

\textbf{Significance Testing:}
A Mann-Whitney U test was conducted to compare the trust between the \Free{} and \Paid{} groups. The results showed no statistically significant difference between the groups for the general trust (U = 1072, p = .7012) and the security trust (U = 1060, p = .7782).

\paragraph{Qualitative Analysis.}
\textbf{Reasons for Trust:}
Forty-one participants reported trusting Gemini’s suggestions because they could assess them using their own knowledge or by cross-referencing with other resources (e.g., \contextquote[F05]{F05}{With my own knowledge [...], I could verify the suggestions [...]. Since they align and make sense, and sometimes can be verified by other sites (SO, API doc, etc.), I trust those suggestions}).
Seventeen participants expressed trust in Gemini’s suggestions because they were mostly correct or worked as expected. 
Twelve participants trusted Gemini’s suggestions due to its informative explanations (e.g., \contextquote[P57]{P57}{I also trusted the code because Gemini explained the code properly and added comments to the code it generated}).

Trust in Gemini's suggestions was linked to its training data, as noted by 10 participants 
(e.g., \contextquote[P14]{P14}{they are based on extensive knowledge and experience}). 
6 participants (P09, P44, F02, F23, F35, F40) trusted because of the company Google behind Gemini (e.g., \contextquote[F23]{F23}{the fact that it was developed by Google as a reputable company helped me trust the suggestions}).
Trust was also attributed to references by 4 participants  (F13, F39, F59, F64),
e.g., \contextquote[F39]{F39}{All the answers by Gemini were referenced. [...] That is why I think Gemini gives trustable results than other LLMs}, \contextquote[F64]{F64}{providing links to official documentation and well-regarded tutorials [...] was particularly important for implementing security [...]}.

Ten participants reported trusting Gemini specifically for security-related suggestions. For example, P27 highlighted Gemini's ability to catch security mistakes: \contextquote[P27]{P27}{giving it a code for, sort of, review provides a layer of security. Basically, it's another pair of eyes that might notice something you've missed or even doesn't know}.

Interestingly, 2 participants (P07, P55) mentioned trusting the paid version of Gemini more: %
\contextquote[P07]{P07}{As its was the premium version, so i believe the resource behind it must be best available},%
\contextquote[P55]{P55}{Looked secure to me, moreover, it is their paid version, so it must generate good code, right?}. Apart from these two participants, there were no instances where trust differed between the free and the paid version, supporting the results of the quantitative analysis.

\textbf{Reasons for Mistrust:}
Twenty-four participants reported that they assessed Gemini’s suggestions using their knowledge or other resources, as they did not trust the AI's suggestions (e.g., \contextquote[P02]{P02}{I googled what Gemini told me just to make sure I'm doing the right thing},  \contextquote[P26]{P26}{It wasn't blind trust, I validated the code using my own knowledge}). 
Similar to working suggestions, non-working suggestions led to mistrust, as mentioned by 26 participants.
Seven participants (F05, F26, F33, F50,  P09, P37, P51) mistrusted Gemini due to its training data (e.g., \contextquote[P51]{P51}{It's totally dependent on humans for training and bad data might be supplied}).
Participants also cited outdated suggestions (P04, P33, P44, P46, P48, F64), required modifications (P13, P15, P33, P5, F04, F25, F61) and misinterpretation of their requirements (P14, P18, P28, P44, P50, P62, F22) (e.g., \contextquote[P18]{P18}{sometimes [Gemini] loses context or ignores instructions previously provided, causes trust issues}).

Twelve participants in \Paid{} and 10 in \Free{} expressed mistrust in Gemini's security-related suggestions (e.g., \contextquote[P09]{P09}{It could scrape the internet and give me some security flaws}). 5 participants (P16, P18, P26, F02, F34) mentioned mistrust because of possible hallucinations (e.g., \contextquote[P16]{P16}{AI Assistants can hallucinate often [...]. This is even more critical in environments where security is paramount}).

\begin{summary}{RQ 4 -- Summary}

Our findings revealed no significant differences in general or security-related trust between Gemini's free and paid versions. Participants reported trusting Gemini's suggestions primarily because they could verify them through their own knowledge or external resources, with some also trusting the AI for its clear explanations, quantity, and quality of training data, or because it met their expectations. Trust was further bolstered by Gemini’s security-related feedback and Google's reputation. 
However, mistrust arose from concerns over incorrect or outdated suggestions, security risks, and the AI’s tendency to ignore context or hallucinate. 
\end{summary}

\section{Discussion and Recommendations}

Our results indicate that Gemini cannot compensate for the depth of expertise that experienced developers bring to secure software development. This has direct implications for an industry already struggling with a persistent shortage of skilled professionals, particularly in security-critical domains. Although many organisations have turned to AI to address this shortage, sometimes even replacing junior developer roles~\cite{Demandfo65:online}, our results suggest that practical programming experience remains essential, especially for producing secure code. While AI assistance may offer value to novice developers, such as learning new concepts or technologies~\cite{kazemitabaar_studying_2023}, it seems not to eliminate the need for human expertise. If organizations increasingly substitute low-experience developers with AI support, junior practitioners will lose vital opportunities to build foundational skills, and the long-term availability of experienced professionals may decline even further, undermining the overall security posture of software ecosystems. 
Based on our findings, we strongly caution against viewing AI tools as substitutes for genuine developer expertise.

\subsection{Implications for Companies}
\textbf{Security Through Programming Experience:}
We found that Gemini did not significantly improve security scores and secure software development could not be fully substituted by Gemini support.
Our analysis showed that developers with more general programming %
experience tend to produce code with significantly higher security scores, supporting the findings of Acar et al.~\cite{acar_security_2017}, who also found a positive impact of self-reported years of experience on security. 
This correlation could occur because more experienced developers might have had more time and workload to focus on security practices. %
Effective workload management could enhance developer satisfaction and code security~\cite{dohmke2023sea,ng2024harnessing, goncales_measuring_2019, muller_measuring_2015}.

\textbf{AI Does Not Eliminate the Need for Security Measures:}

While Gemini could replace programming experience, it can assist
developers, particularly in security-critical implementations such as password storage. Despite these benefits, our findings indicate that the security scores of code written with Gemini assistance was comparable to that of code written without such assistance, supporting the findings of prior studies~\cite{asare_user-centered_2024, sandoval_lost_2022}. Consequently, organizations should not rely on AI tools as a substitute for security processes within the software development lifecycle~\cite{assal_security_2018}. Security testing, code analysis and security code reviews remain essential regardless of whether AI assistance is used.

\subsection{Implications for Developers}
\textbf{Don't Rely Solely on AI for Security:}
Developers should be cautious about relying on AI tools such as Gemini for enhanced security. In the context of our study, we did not observe significant differences in trust between the free and the paid versions, however, 2 participants mentioned trusting Gemini Advanced more because it is the paid version.
Instead of depending on AI-generated code for security measures, developers might use AI as an extra layer of support, particularly for identifying potential vulnerabilities or addressing security measures that may have been overlooked. 
As our participants frequently mentioned errors in security-related suggestions or newly introduced vulnerabilities from Gemini, it is critical to validate AI’s suggestions with trusted resources, such as OWASP~\cite{Aboutthe44:online} or NIST~\cite{Cybersec32:online}. 

\textbf{Security Overconfidence:}
Our study revealed an unexpected trend indicating that participants with self-reported prior security experience tended to write less secure code. A possible explanation could be that participants who feel more confident in their security skills might believe they are less likely to make mistakes, leading them to pay less attention to security and overlook vulnerabilities. 
Code reviews focused on security~\cite{braz_less_2022}, along with the use of static analysis tools, can help identify overlooked vulnerabilities~\cite{tahaei_security_2021}.

\subsection{Implications for AI Practitioners}
\textbf{Vulnerabilities:}
While Asare et al.~\cite{asare_user-centered_2024} found no significant impact of AI assistance on specific vulnerability types, our findings suggest a nuanced effect.
When considering the individual vulnerabilities, Gemini assistance led to approximately 21.7\% fewer vulnerabilities in password storage and 10.5\% fewer in SQL vulnerabilities. However, participants still faced challenges, such as Gemini suggesting outdated libraries or omitting hashing recommendations. XSS vulnerabilities were improved by only 3.5\% with the use of Gemini. CSRF and improper input validation seemed almost not to be influenced by Gemini usage. This suggests that AI tool designers might require to address specific vulnerabilities.
For instance, AI training could be improved by running static analysis checks before including code into the training data~\cite{perry_users_2022}, including only the latest versions of security libraries to avoid outdated suggestions.

\textbf{When Paying More Does Not Mean More Secure:}
While our participants reported that Gemini helped them to implement security measures, no significant improvement in security scores was observed between the free and paid versions. 
Their responses showed that Gemini frequently overlooked security concerns without explicit prompting. AI-generated code intended to address security often contained errors, making it unsuitable for implementation. These findings reflected that developer confidence in Gemini's paid and free versions did not differ. 
On the one hand, this could indicate that the paid version's benefits are not being communicated effectively or that these benefits were not apparent when the participants used the paid version of Gemini.
On the other hand, these findings also suggest that organizations and individual developers might not be required investing in costly paid versions of AI tools if free versions offer comparable security.

\section{Conclusion}
To explore the impact of AI tools such as Gemini on software security, we conducted a study with 159 developers recruited through the freelancer platform Upwork. 
Our findings showed that developers using Gemini did not produce code that was significantly more secure.
We did not observe significant differences in the security or user trust between the free and paid versions. However, programming experience improved code security significantly. 
While developers with no security experience demonstrated enhanced code security when supported by Gemini,
we advise against relying solely on AI tools and recommend using them as a supplementary layer.

\begin{acks}
This work was funded by the Deutsche Forschungsgemeinschaft (DFG, German Research Foundation) under Germany’s Excellence Strategy - EXC 2092 CASA - 390781972.
\end{acks}

\bibliographystyle{ACM-Reference-Format}
\bibliography{literature-base}


\begin{thebibliography}{84}


\ifx \showCODEN    \undefined \def \showCODEN     #1{\unskip}     \fi
\ifx \showISBNx    \undefined \def \showISBNx     #1{\unskip}     \fi
\ifx \showISBNxiii \undefined \def \showISBNxiii  #1{\unskip}     \fi
\ifx \showISSN     \undefined \def \showISSN      #1{\unskip}     \fi
\ifx \showLCCN     \undefined \def \showLCCN      #1{\unskip}     \fi
\ifx \shownote     \undefined \def \shownote      #1{#1}          \fi
\ifx \showarticletitle \undefined \def \showarticletitle #1{#1}   \fi
\ifx \showURL      \undefined \def \showURL       {\relax}        \fi
\providecommand\bibfield[2]{#2}
\providecommand\bibinfo[2]{#2}
\providecommand\natexlab[1]{#1}
\providecommand\showeprint[2][]{arXiv:#2}

\bibitem[Acar et~al\mbox{.}(2016)]%
        {acar_you_2016}
\bibfield{author}{\bibinfo{person}{Yasemin Acar}, \bibinfo{person}{Michael Backes}, \bibinfo{person}{Sascha Fahl}, \bibinfo{person}{Doowon Kim}, \bibinfo{person}{Michelle~L. Mazurek}, {and} \bibinfo{person}{Christian Stransky}.} \bibinfo{year}{2016}\natexlab{}.
\newblock \showarticletitle{You Get Where You're Looking for: The Impact of Information Sources on Code Security}. In \bibinfo{booktitle}{\emph{2016 {IEEE} Symposium on Security and Privacy ({SP})}} (San Jose, {CA}, 2016-05). \bibinfo{publisher}{{IEEE}}, \bibinfo{address}{San Francisco, CA, US}, \bibinfo{pages}{289--305}.
\newblock
\showISBNx{978-1-5090-0824-7}
\href{https://doi.org/10.1109/SP.2016.25}{doi:\nolinkurl{10.1109/SP.2016.25}}


\bibitem[Acar et~al\mbox{.}(2017)]%
        {acar_security_2017}
\bibfield{author}{\bibinfo{person}{Yasemin Acar}, \bibinfo{person}{Christian Stransky}, \bibinfo{person}{Dominik Wermke}, \bibinfo{person}{Michelle~L. Mazurek}, {and} \bibinfo{person}{Sascha Fahl}.} \bibinfo{year}{2017}\natexlab{}.
\newblock \showarticletitle{Security developer studies with github users: exploring a convenience sample}. In \bibinfo{booktitle}{\emph{Proceedings of the Thirteenth USENIX Conference on Usable Privacy and Security}} (Santa Clara, CA, USA) \emph{(\bibinfo{series}{SOUPS '17})}. \bibinfo{publisher}{USENIX Association}, \bibinfo{address}{USA}, \bibinfo{pages}{81–95}.
\newblock
\showISBNx{9781931971393}


\bibitem[Asare et~al\mbox{.}(2024)]%
        {asare_user-centered_2024}
\bibfield{author}{\bibinfo{person}{Owura Asare}, \bibinfo{person}{Meiyappan Nagappan}, {and} \bibinfo{person}{N. Asokan}.} \bibinfo{year}{2024}\natexlab{}.
\newblock \showarticletitle{A User-centered Security Evaluation of Copilot}. In \bibinfo{booktitle}{\emph{Proceedings of the IEEE/ACM 46th International Conference on Software Engineering}} (Lisbon, Portugal) \emph{(\bibinfo{series}{ICSE '24})}. \bibinfo{publisher}{Association for Computing Machinery}, \bibinfo{address}{New York, NY, USA}, Article \bibinfo{articleno}{158}, \bibinfo{numpages}{11}~pages.
\newblock
\showISBNx{9798400702174}
\href{https://doi.org/10.1145/3597503.3639154}{doi:\nolinkurl{10.1145/3597503.3639154}}


\bibitem[Assal and Chiasson(2018)]%
        {assal_security_2018}
\bibfield{author}{\bibinfo{person}{Hala Assal} {and} \bibinfo{person}{Sonia Chiasson}.} \bibinfo{year}{2018}\natexlab{}.
\newblock \showarticletitle{Security in the Software Development Lifecycle}. In \bibinfo{booktitle}{\emph{Fourteenth Symposium on Usable Privacy and Security (SOUPS 2018)}}. \bibinfo{publisher}{USENIX Association}, \bibinfo{address}{Baltimore, MD}, \bibinfo{pages}{281--296}.
\newblock
\showISBNx{978-1-939133-10-6}
\urldef\tempurl%
\url{https://www.usenix.org/conference/soups2018/presentation/assal}
\showURL{%
\tempurl}


\bibitem[Austin et~al\mbox{.}(2021)]%
        {austin_program_2021}
\bibfield{author}{\bibinfo{person}{Jacob Austin}, \bibinfo{person}{Augustus Odena}, \bibinfo{person}{Maxwell Nye}, \bibinfo{person}{Maarten Bosma}, \bibinfo{person}{Henryk Michalewski}, \bibinfo{person}{David Dohan}, \bibinfo{person}{Ellen Jiang}, \bibinfo{person}{Carrie Cai}, \bibinfo{person}{Michael Terry}, \bibinfo{person}{Quoc Le}, {and} \bibinfo{person}{Charles Sutton}.} \bibinfo{year}{2021}\natexlab{}.
\newblock \bibinfo{title}{Program Synthesis with Large Language Models}.
\newblock
\showeprint[arxiv]{2108.07732 [cs]}
\urldef\tempurl%
\url{http://arxiv.org/abs/2108.07732}
\showURL{%
\tempurl}


\bibitem[Bamberger et~al\mbox{.}(2020)]%
        {bamberger2020can}
\bibfield{author}{\bibinfo{person}{Kenneth~A Bamberger}, \bibinfo{person}{Serge Egelman}, \bibinfo{person}{Catherine Han}, \bibinfo{person}{Amit~Elazari Bar~On}, {and} \bibinfo{person}{Irwin Reyes}.} \bibinfo{year}{2020}\natexlab{}.
\newblock \showarticletitle{Can you pay for privacy? consumer expectations and the behavior of free and paid apps}.
\newblock \bibinfo{journal}{\emph{Berkeley Tech. LJ}}  \bibinfo{volume}{35} (\bibinfo{year}{2020}), \bibinfo{pages}{327}.
\newblock


\bibitem[Biryukov et~al\mbox{.}(2015)]%
        {biryukov_argon2_2015}
\bibfield{author}{\bibinfo{person}{Alex Biryukov}, \bibinfo{person}{Daniel Dinu}, {and} \bibinfo{person}{Dmitry Khovratovich}.} \bibinfo{year}{2015}\natexlab{}.
\newblock \bibinfo{title}{Argon2: the memory-hard function for password hashing and other applications}.
\newblock
\urldef\tempurl%
\url{https://www.password-hashing.net/argon2-specs.pdf}
\showURL{%
\tempurl}
\newblock
\shownote{[Online; accessed 2025-12-11]}.


\bibitem[Boyatzis(1998)]%
        {boyatzis_transforming_1998}
\bibfield{author}{\bibinfo{person}{Richard~E. Boyatzis}.} \bibinfo{year}{1998}\natexlab{}.
\newblock \bibinfo{booktitle}{\emph{Transforming Qualitative Information {\textbar} {SAGE} Publications Inc}}.
\newblock
\urldef\tempurl%
\url{https://us.sagepub.com/en-us/nam/transforming-qualitative-information/book7714}
\showURL{%
\tempurl}


\bibitem[Braun and Clarke(2024)]%
        {Gotquest27:online}
\bibfield{author}{\bibinfo{person}{Virginia Braun} {and} \bibinfo{person}{Victoria Clarke}.} \bibinfo{year}{2024}\natexlab{}.
\newblock \bibinfo{title}{Got questions about Thematic Analysis? We have prepared some answers to common ones.}
\newblock \bibinfo{howpublished}{\url{https://www.thematicanalysis.net/faqs/}}.
\newblock
\newblock
\shownote{[Online; accessed 2025-12-05]}.


\bibitem[Braun et~al\mbox{.}(2022)]%
        {braun_starting_2022}
\bibfield{author}{\bibinfo{person}{Virginia Braun}, \bibinfo{person}{Victoria Clarke}, {and} \bibinfo{person}{Nikki Hayfield}.} \bibinfo{year}{2022}\natexlab{}.
\newblock \showarticletitle{‘A starting point for your journey, not a map’: Nikki Hayfield in conversation with Virginia Braun and Victoria Clarke about thematic analysis}.
\newblock \bibinfo{journal}{\emph{Qualitative Research in Psychology}} \bibinfo{volume}{19}, \bibinfo{number}{2} (\bibinfo{year}{2022}), \bibinfo{pages}{424--445}.
\newblock
\showISSN{1478-0887}
\href{https://doi.org/10.1080/14780887.2019.1670765}{doi:\nolinkurl{10.1080/14780887.2019.1670765}}


\bibitem[Braz et~al\mbox{.}(2022)]%
        {braz_less_2022}
\bibfield{author}{\bibinfo{person}{Larissa Braz}, \bibinfo{person}{Christian Aeberhard}, \bibinfo{person}{Gul Calikli}, {and} \bibinfo{person}{Alberto Bacchelli}.} \bibinfo{year}{2022}\natexlab{}.
\newblock \showarticletitle{{ Less is More: Supporting Developers in Vulnerability Detection during Code Review }}. In \bibinfo{booktitle}{\emph{2022 IEEE/ACM 44th International Conference on Software Engineering (ICSE)}}. \bibinfo{publisher}{IEEE Computer Society}, \bibinfo{address}{Los Alamitos, CA, USA}, \bibinfo{pages}{1317--1329}.
\newblock
\href{https://doi.org/10.1145/3510003.3511560}{doi:\nolinkurl{10.1145/3510003.3511560}}


\bibitem[Brooke(1996)]%
        {brooke1996sus}
\bibfield{author}{\bibinfo{person}{J Brooke}.} \bibinfo{year}{1996}\natexlab{}.
\newblock \showarticletitle{SUS: A “quick and dirty” Usability Scale}.
\newblock \bibinfo{journal}{\emph{Usability evaluation in industry}} \bibinfo{volume}{189}, \bibinfo{number}{194} (\bibinfo{year}{1996}), \bibinfo{pages}{4--7}.
\newblock


\bibitem[BV(2024)]%
        {TIOBEInd0:online}
\bibfield{author}{\bibinfo{person}{TIOBE~Software BV}.} \bibinfo{year}{2024}\natexlab{}.
\newblock \bibinfo{title}{TIOBE Index - TIOBE}.
\newblock \bibinfo{howpublished}{\url{https://www.tiobe.com/tiobe-index/}}.
\newblock
\newblock
\shownote{[Online; accessed 2025-12-05]}.


\bibitem[Chandra(2024)]%
        {Chandra:online}
\bibfield{author}{\bibinfo{person}{Satish Chandra}.} \bibinfo{year}{2024}\natexlab{}.
\newblock \bibinfo{title}{Progress of AI-based assistance for software engineering in Google’s internal tooling and our projections for the future.}
\newblock
\urldef\tempurl%
\url{https://research.google/blog/ai-in-software-engineering-at-google-progress-and-the-path-ahead/}
\showURL{%
\tempurl}
\newblock
\shownote{[Online; accessed 2025-12-05]}.


\bibitem[Chen et~al\mbox{.}(2021)]%
        {chen_evaluating_2021}
\bibfield{author}{\bibinfo{person}{Mark Chen}, \bibinfo{person}{Jerry Tworek}, \bibinfo{person}{Heewoo Jun}, \bibinfo{person}{Qiming Yuan}, \bibinfo{person}{Henrique Ponde de~Oliveira Pinto}, \bibinfo{person}{Jared Kaplan}, \bibinfo{person}{Harri Edwards}, \bibinfo{person}{Yuri Burda}, \bibinfo{person}{Nicholas Joseph}, \bibinfo{person}{Greg Brockman}, \bibinfo{person}{Alex Ray}, \bibinfo{person}{Raul Puri}, \bibinfo{person}{Gretchen Krueger}, \bibinfo{person}{Michael Petrov}, \bibinfo{person}{Heidy Khlaaf}, \bibinfo{person}{Girish Sastry}, \bibinfo{person}{Pamela Mishkin}, \bibinfo{person}{Brooke Chan}, \bibinfo{person}{Scott Gray}, \bibinfo{person}{Nick Ryder}, \bibinfo{person}{Mikhail Pavlov}, \bibinfo{person}{Alethea Power}, \bibinfo{person}{Lukasz Kaiser}, \bibinfo{person}{Mohammad Bavarian}, \bibinfo{person}{Clemens Winter}, \bibinfo{person}{Philippe Tillet}, \bibinfo{person}{Felipe~Petroski Such}, \bibinfo{person}{Dave Cummings}, \bibinfo{person}{Matthias Plappert}, \bibinfo{person}{Fotios Chantzis},
  \bibinfo{person}{Elizabeth Barnes}, \bibinfo{person}{Ariel Herbert-Voss}, \bibinfo{person}{William~Hebgen Guss}, \bibinfo{person}{Alex Nichol}, \bibinfo{person}{Alex Paino}, \bibinfo{person}{Nikolas Tezak}, \bibinfo{person}{Jie Tang}, \bibinfo{person}{Igor Babuschkin}, \bibinfo{person}{Suchir Balaji}, \bibinfo{person}{Shantanu Jain}, \bibinfo{person}{William Saunders}, \bibinfo{person}{Christopher Hesse}, \bibinfo{person}{Andrew~N. Carr}, \bibinfo{person}{Jan Leike}, \bibinfo{person}{Josh Achiam}, \bibinfo{person}{Vedant Misra}, \bibinfo{person}{Evan Morikawa}, \bibinfo{person}{Alec Radford}, \bibinfo{person}{Matthew Knight}, \bibinfo{person}{Miles Brundage}, \bibinfo{person}{Mira Murati}, \bibinfo{person}{Katie Mayer}, \bibinfo{person}{Peter Welinder}, \bibinfo{person}{Bob {McGrew}}, \bibinfo{person}{Dario Amodei}, \bibinfo{person}{Sam {McCandlish}}, \bibinfo{person}{Ilya Sutskever}, {and} \bibinfo{person}{Wojciech Zaremba}.} \bibinfo{year}{2021}\natexlab{}.
\newblock \bibinfo{title}{Evaluating Large Language Models Trained on Code}.
\newblock
\showeprint[arxiv]{2107.03374 [cs]}
\urldef\tempurl%
\url{http://arxiv.org/abs/2107.03374}
\showURL{%
\tempurl}


\bibitem[Cohen(1988)]%
        {cohen_statistical_1988}
\bibfield{author}{\bibinfo{person}{Jacob Cohen}.} \bibinfo{year}{1988}\natexlab{}.
\newblock \bibinfo{booktitle}{\emph{Statistical {Power} {Analysis} for the {Behavioral} {Sciences}} (\bibinfo{edition}{2} ed.)}.
\newblock \bibinfo{publisher}{Routledge}, \bibinfo{address}{New York}.
\newblock
\showISBNx{978-0-203-77158-7}
\href{https://doi.org/10.4324/9780203771587}{doi:\nolinkurl{10.4324/9780203771587}}


\bibitem[Commission(2024)]%
        {meta_judgment}
\bibfield{author}{\bibinfo{person}{Data~Protection Commission}.} \bibinfo{year}{2024}\natexlab{}.
\newblock \bibinfo{title}{Irish Data Protection Commission fines Meta Ireland €91 million}.
\newblock \bibinfo{howpublished}{\url{https://www.dataprotection.ie/en/news-media/press-releases/DPC-announces-91-million-fine-of-Meta}}.
\newblock
\newblock
\shownote{[Online; accessed 2025-12-05]}.


\bibitem[Dohmke et~al\mbox{.}(2023)]%
        {dohmke2023sea}
\bibfield{author}{\bibinfo{person}{Thomas Dohmke}, \bibinfo{person}{Marco Iansiti}, {and} \bibinfo{person}{Greg Richards}.} \bibinfo{year}{2023}\natexlab{}.
\newblock \bibinfo{title}{Sea change in software development: Economic and productivity analysis of the ai-powered developer lifecycle}.
\newblock


\bibitem[Du et~al\mbox{.}(2024)]%
        {du_evaluating_2024}
\bibfield{author}{\bibinfo{person}{Xueying Du}, \bibinfo{person}{Mingwei Liu}, \bibinfo{person}{Kaixin Wang}, \bibinfo{person}{Hanlin Wang}, \bibinfo{person}{Junwei Liu}, \bibinfo{person}{Yixuan Chen}, \bibinfo{person}{Jiayi Feng}, \bibinfo{person}{Chaofeng Sha}, \bibinfo{person}{Xin Peng}, {and} \bibinfo{person}{Yiling Lou}.} \bibinfo{year}{2024}\natexlab{}.
\newblock \showarticletitle{Evaluating Large Language Models in Class-Level Code Generation}. In \bibinfo{booktitle}{\emph{Proceedings of the {IEEE}/{ACM} 46th International Conference on Software Engineering}} (New York, {NY}, {USA}, 2024-04-12) \emph{(\bibinfo{series}{{ICSE} '24})}. \bibinfo{publisher}{Association for Computing Machinery}, \bibinfo{address}{Lisbon, Portugal}, \bibinfo{pages}{1--13}.
\newblock
\showISBNx{9798400702174}
\href{https://doi.org/10.1145/3597503.3639219}{doi:\nolinkurl{10.1145/3597503.3639219}}


\bibitem[Friedmann(2025)]%
        {Demandfo65:online}
\bibfield{author}{\bibinfo{person}{By~Carl Friedmann}.} \bibinfo{year}{2025}\natexlab{}.
\newblock \bibinfo{title}{Demand for junior developers softens as AI takes over | CIO}.
\newblock
\urldef\tempurl%
\url{https://www.cio.com/article/4062024/demand-for-junior-developers-softens-as-ai-takes-over.html}
\showURL{%
\tempurl}
\newblock
\shownote{[Online; accessed 2025-12-10]}.


\bibitem[GitHub(2024)]%
        {GitHubCo85:online}
\bibfield{author}{\bibinfo{person}{Inc. GitHub}.} \bibinfo{year}{2024}\natexlab{}.
\newblock \bibinfo{title}{GitHub Copilot · Your AI pair programmer}.
\newblock \bibinfo{howpublished}{\url{https://github.com/features/copilot}}.
\newblock
\newblock
\shownote{[Online; accessed 2025-12-05]}.


\bibitem[{GitHub, Inc.}(2024)]%
        {Surveyre91:online}
\bibfield{author}{\bibinfo{person}{{GitHub, Inc.}}} \bibinfo{year}{2024}\natexlab{}.
\newblock \bibinfo{title}{Survey reveals AI’s impact on the developer experience - The GitHub Blog}.
\newblock \bibinfo{howpublished}{\url{https://github.blog/news-insights/research/survey-reveals-ais-impact-on-the-developer-experience/}}.
\newblock
\newblock
\shownote{[Online; accessed 2025-12-05]}.


\bibitem[Gonçales et~al\mbox{.}(2019)]%
        {goncales_measuring_2019}
\bibfield{author}{\bibinfo{person}{Lucian Gonçales}, \bibinfo{person}{Kleinner Farias}, \bibinfo{person}{Bruno da Silva}, {and} \bibinfo{person}{Jonathan Fessler}.} \bibinfo{year}{2019}\natexlab{}.
\newblock \showarticletitle{Measuring the Cognitive Load of Software Developers: A Systematic Mapping Study}. In \bibinfo{booktitle}{\emph{2019 {IEEE}/{ACM} 27th International Conference on Program Comprehension ({ICPC})}}. \bibinfo{publisher}{IEEE/ACM}, \bibinfo{address}{Montreal, QC, Canada}, \bibinfo{pages}{42--52}.
\newblock
\href{https://doi.org/10.1109/ICPC.2019.00018}{doi:\nolinkurl{10.1109/ICPC.2019.00018}}
\newblock
\shownote{{ISSN}: 2643-7171}.


\bibitem[Google(2024)]%
        {GoogleBa3:online}
\bibfield{author}{\bibinfo{person}{Google}.} \bibinfo{year}{2024}\natexlab{}.
\newblock \bibinfo{title}{Bard becomes Gemini: Try Ultra 1.0 and a new mobile app today}.
\newblock \bibinfo{howpublished}{\url{https://blog.google/products/gemini/bard-gemini-advanced-app/}}.
\newblock
\newblock
\shownote{[Online; accessed 2025-12-05]}.


\bibitem[Google(2025)]%
        {Gemini57:online}
\bibfield{author}{\bibinfo{person}{Google}.} \bibinfo{year}{2025}\natexlab{}.
\newblock \bibinfo{title}{Gemini}.
\newblock \bibinfo{howpublished}{\url{https://gemini.google.com/app}}.
\newblock
\newblock
\shownote{[Online; accessed 2025-12-05]}.


\bibitem[Graefe(2025)]%
        {Howsoftw68:online}
\bibfield{author}{\bibinfo{person}{Jason Graefe}.} \bibinfo{year}{2025}\natexlab{}.
\newblock \bibinfo{title}{How software development companies are paving the way for AI transformation}.
\newblock
\urldef\tempurl%
\url{https://partner.microsoft.com/en-kw/blog/article/generative-ai-impact-for-partners}
\showURL{%
\tempurl}
\newblock
\shownote{[Online; accessed 2025-12-05]}.


\bibitem[Hamer et~al\mbox{.}(2024)]%
        {hamer_just_2024}
\bibfield{author}{\bibinfo{person}{Sivana Hamer}, \bibinfo{person}{Marcelo d’Amorim}, {and} \bibinfo{person}{Laurie Williams}.} \bibinfo{year}{2024}\natexlab{}.
\newblock \showarticletitle{Just another copy and paste? Comparing the security vulnerabilities of ChatGPT generated code and StackOverflow answers}. In \bibinfo{booktitle}{\emph{2024 IEEE Security and Privacy Workshops (SPW)}}. \bibinfo{publisher}{IEEE}, \bibinfo{address}{San Francisco, CA, US}, \bibinfo{pages}{87--94}.
\newblock
\href{https://doi.org/10.1109/SPW63631.2024.00014}{doi:\nolinkurl{10.1109/SPW63631.2024.00014}}


\bibitem[Hart and Staveland(1988)]%
        {hart1988development}
\bibfield{author}{\bibinfo{person}{Sandra~G Hart} {and} \bibinfo{person}{Lowell~E Staveland}.} \bibinfo{year}{1988}\natexlab{}.
\newblock \showarticletitle{Development of NASA-TLX (Task Load Index): Results of empirical and theoretical research}.
\newblock In \bibinfo{booktitle}{\emph{Advances in psychology}}. Vol.~\bibinfo{volume}{52}. \bibinfo{publisher}{Elsevier}, \bibinfo{pages}{139--183}.
\newblock


\bibitem[Hassan(2026)]%
        {hassan2026agentic}
\bibfield{author}{\bibinfo{person}{Ahmed~E. Hassan}.} \bibinfo{year}{2026}\natexlab{}.
\newblock \bibinfo{booktitle}{\emph{Agentic Software Engineering: Building Trustworthy Software with Stochastic Teammates at Unprecedented Scale} (\bibinfo{edition}{first edition, v0.3} ed.)}.
\newblock


\bibitem[Hatzivasilis et~al\mbox{.}(2015)]%
        {hatzivasilis_password_2015}
\bibfield{author}{\bibinfo{person}{George Hatzivasilis}, \bibinfo{person}{Ioannis Papaefstathiou}, {and} \bibinfo{person}{Charalampos Manifavas}.} \bibinfo{year}{2015}\natexlab{}.
\newblock \bibinfo{title}{Password Hashing Competition - Survey and Benchmark}.
\newblock \bibinfo{howpublished}{Cryptology {ePrint} Archive, Paper 2015/265}.
\newblock
\urldef\tempurl%
\url{https://eprint.iacr.org/2015/265}
\showURL{%
\tempurl}


\bibitem[He and Vechev(2023)]%
        {he_large_2023}
\bibfield{author}{\bibinfo{person}{Jingxuan He} {and} \bibinfo{person}{Martin Vechev}.} \bibinfo{year}{2023}\natexlab{}.
\newblock \showarticletitle{Large Language Models for Code: Security Hardening and Adversarial Testing}. In \bibinfo{booktitle}{\emph{Proceedings of the 2023 {ACM} {SIGSAC} Conference on Computer and Communications Security}} (2023-11-15). \bibinfo{publisher}{{ACM}}, \bibinfo{address}{Copenhagen Denmark}, \bibinfo{pages}{1865--1879}.
\newblock
\showISBNx{9798400700507}
\href{https://doi.org/10.1145/3576915.3623175}{doi:\nolinkurl{10.1145/3576915.3623175}}


\bibitem[Hou and Ji(2024)]%
        {hou_systematic_2024}
\bibfield{author}{\bibinfo{person}{Wenpin Hou} {and} \bibinfo{person}{Zhicheng Ji}.} \bibinfo{year}{2024}\natexlab{}.
\newblock \bibinfo{title}{Comparing Large Language Models and Human Programmers for Generating Programming Code}.
\newblock
\showISSN{2198-3844}
\href{https://doi.org/10.1002/advs.202412279}{doi:\nolinkurl{10.1002/advs.202412279}}


\bibitem[Inc.(2024)]%
        {UpworkTh45:online}
\bibfield{author}{\bibinfo{person}{Upwork®~Global Inc.}} \bibinfo{year}{2024}\natexlab{}.
\newblock \bibinfo{title}{Upwork | The World’s Work Marketplace}.
\newblock \bibinfo{howpublished}{\url{https://www.upwork.com/}}.
\newblock
\newblock
\shownote{[Online; accessed 2025-12-05]}.


\bibitem[Kaur et~al\mbox{.}(2022)]%
        {kaur_where_2022}
\bibfield{author}{\bibinfo{person}{Harjot Kaur}, \bibinfo{person}{Sabrina Amft}, \bibinfo{person}{Daniel Votipka}, \bibinfo{person}{Yasemin Acar}, {and} \bibinfo{person}{Sascha Fahl}.} \bibinfo{year}{2022}\natexlab{}.
\newblock \showarticletitle{Where to Recruit for Security Development Studies: Comparing Six Software Developer Samples}. In \bibinfo{booktitle}{\emph{31st USENIX Security Symposium (USENIX Security 22)}}. \bibinfo{publisher}{USENIX Association}, \bibinfo{address}{Boston, MA}, \bibinfo{pages}{4041--4058}.
\newblock
\showISBNx{978-1-939133-31-1}
\urldef\tempurl%
\url{https://www.usenix.org/conference/usenixsecurity22/presentation/kaur}
\showURL{%
\tempurl}


\bibitem[Kazemitabaar et~al\mbox{.}(2023)]%
        {kazemitabaar_studying_2023}
\bibfield{author}{\bibinfo{person}{Majeed Kazemitabaar}, \bibinfo{person}{Justin Chow}, \bibinfo{person}{Carl Ka~To Ma}, \bibinfo{person}{Barbara~J. Ericson}, \bibinfo{person}{David Weintrop}, {and} \bibinfo{person}{Tovi Grossman}.} \bibinfo{year}{2023}\natexlab{}.
\newblock \showarticletitle{Studying the effect of {AI} Code Generators on Supporting Novice Learners in Introductory Programming}. In \bibinfo{booktitle}{\emph{Proceedings of the 2023 {CHI} Conference on Human Factors in Computing Systems}} (2023-04-19). \bibinfo{publisher}{{ACM}}, \bibinfo{address}{Hamburg Germany}, \bibinfo{pages}{1--23}.
\newblock
\showISBNx{978-1-4503-9421-5}
\href{https://doi.org/10.1145/3544548.3580919}{doi:\nolinkurl{10.1145/3544548.3580919}}


\bibitem[Kyle~Daigle(2024)]%
        {SurveyTh78:online}
\bibfield{author}{\bibinfo{person}{GitHub~Staff Kyle~Daigle}.} \bibinfo{year}{2024}\natexlab{}.
\newblock \bibinfo{title}{Survey: The AI wave continues to grow on software development teams - The GitHub Blog}.
\newblock
\urldef\tempurl%
\url{https://github.blog/news-insights/research/survey-ai-wave-grows/}
\showURL{%
\tempurl}
\newblock
\shownote{[Online; accessed 2025-12-05]}.


\bibitem[Li et~al\mbox{.}(2022)]%
        {li_competition-level_2022}
\bibfield{author}{\bibinfo{person}{Yujia Li}, \bibinfo{person}{David Choi}, \bibinfo{person}{Junyoung Chung}, \bibinfo{person}{Nate Kushman}, \bibinfo{person}{Julian Schrittwieser}, \bibinfo{person}{Rémi Leblond}, \bibinfo{person}{Tom Eccles}, \bibinfo{person}{James Keeling}, \bibinfo{person}{Felix Gimeno}, \bibinfo{person}{Agustin~Dal Lago}, \bibinfo{person}{Thomas Hubert}, \bibinfo{person}{Peter Choy}, \bibinfo{person}{Cyprien de~Masson d'Autume}, \bibinfo{person}{Igor Babuschkin}, \bibinfo{person}{Xinyun Chen}, \bibinfo{person}{Po-Sen Huang}, \bibinfo{person}{Johannes Welbl}, \bibinfo{person}{Sven Gowal}, \bibinfo{person}{Alexey Cherepanov}, \bibinfo{person}{James Molloy}, \bibinfo{person}{Daniel~J. Mankowitz}, \bibinfo{person}{Esme~Sutherland Robson}, \bibinfo{person}{Pushmeet Kohli}, \bibinfo{person}{Nando de Freitas}, \bibinfo{person}{Koray Kavukcuoglu}, {and} \bibinfo{person}{Oriol Vinyals}.} \bibinfo{year}{2022}\natexlab{}.
\newblock \showarticletitle{Competition-Level Code Generation with {AlphaCode}}.
\newblock \bibinfo{journal}{\emph{Science}} \bibinfo{volume}{378}, \bibinfo{number}{6624} (\bibinfo{year}{2022}), \bibinfo{pages}{1092--1097}.
\newblock
\showISSN{0036-8075, 1095-9203}
\showeprint[arxiv]{2203.07814 [cs]}
\href{https://doi.org/10.1126/science.abq1158}{doi:\nolinkurl{10.1126/science.abq1158}}


\bibitem[Liang et~al\mbox{.}(2024)]%
        {liang_large-scale_2024}
\bibfield{author}{\bibinfo{person}{Jenny~T. Liang}, \bibinfo{person}{Chenyang Yang}, {and} \bibinfo{person}{Brad~A. Myers}.} \bibinfo{year}{2024}\natexlab{}.
\newblock \showarticletitle{A Large-Scale Survey on the Usability of {AI} Programming Assistants: Successes and Challenges}. In \bibinfo{booktitle}{\emph{Proceedings of the {IEEE}/{ACM} 46th International Conference on Software Engineering (ICSE)}} (2024-02-06). \bibinfo{publisher}{{ACM}}, \bibinfo{address}{Lisbon Portugal}, \bibinfo{pages}{1--13}.
\newblock
\href{https://doi.org/10.1145/3597503.3608128}{doi:\nolinkurl{10.1145/3597503.3608128}}


\bibitem[Linktree(2024)]%
        {Linkinbi74:online}
\bibfield{author}{\bibinfo{person}{Linktree}.} \bibinfo{year}{2024}\natexlab{}.
\newblock \bibinfo{title}{Link in bio tool: Everything you are, in one simple link | Linktree}.
\newblock \bibinfo{howpublished}{\url{https://linktr.ee/}}.
\newblock
\newblock
\shownote{[Online; accessed 2025-12-05]}.


\bibitem[Mattioli and Malatras(2024)]%
        {mattioli2024foresight}
\bibfield{author}{\bibinfo{person}{Rossella Mattioli} {and} \bibinfo{person}{Apostolos Malatras}.} \bibinfo{year}{2024}\natexlab{}.
\newblock \bibinfo{booktitle}{\emph{Foresight cybersecurity threats for 2030: Update}}.
\newblock
\urldef\tempurl%
\url{https://www.enisa.europa.eu/sites/default/files/2024-11/Cybersecurity%20Threats%20for%202030%20-%20Update%202024%20-%20Executive%20Summary_0.pdf}
\showURL{%
\tempurl}
\newblock
\shownote{[Online; accessed 2025-12-05]}.


\bibitem[Morales(2026)]%
        {morales2026amazon_genai_changes}
\bibfield{author}{\bibinfo{person}{Jowi Morales}.} \bibinfo{year}{2026}\natexlab{}.
\newblock \bibinfo{title}{In wake of outage, Amazon calls upon senior engineers to address issues created by 'Gen-AI assisted changes,' report claims}.
\newblock
\urldef\tempurl%
\url{https://www.tomshardware.com/tech-industry/artificial-intelligence/amazon-calls-engineers-to-address-issues-caused-by-use-of-ai-tools-report-claims-company-says-recent-incidents-had-high-blast-radius-and-were-allegedly-related-to-gen-ai-assisted-changes}
\showURL{%
\tempurl}
\newblock
\shownote{Accessed: 2026-03-13}.


\bibitem[Mousavi et~al\mbox{.}(2024)]%
        {mousavi_investigation_2024}
\bibfield{author}{\bibinfo{person}{Zahra Mousavi}, \bibinfo{person}{Chadni Islam}, \bibinfo{person}{Kristen Moore}, \bibinfo{person}{Alsharif Abuadbba}, {and} \bibinfo{person}{M.~Ali Babar}.} \bibinfo{year}{2024}\natexlab{}.
\newblock \showarticletitle{An Investigation into Misuse of Java Security {APIs} by Large Language Models}. In \bibinfo{booktitle}{\emph{Proceedings of the 19th {ACM} Asia Conference on Computer and Communications Security}} (Singapore Singapore, 2024-07). \bibinfo{publisher}{{ACM}}, \bibinfo{address}{Singapore}, \bibinfo{pages}{1299--1315}.
\newblock
\showISBNx{979-8-4007-0482-6}
\href{https://doi.org/10.1145/3634737.3661134}{doi:\nolinkurl{10.1145/3634737.3661134}}


\bibitem[Muller(2015)]%
        {muller_measuring_2015}
\bibfield{author}{\bibinfo{person}{Sebastian~C. Muller}.} \bibinfo{year}{2015}\natexlab{}.
\newblock \showarticletitle{Measuring Software Developers' Perceived Difficulty with Biometric Sensors}. In \bibinfo{booktitle}{\emph{2015 {IEEE}/{ACM} 37th {IEEE} International Conference on Software Engineering}} (2015-05). \bibinfo{publisher}{{IEEE}}, \bibinfo{address}{Florence, Italy}, \bibinfo{pages}{887--890}.
\newblock
\showISBNx{978-1-4799-1934-5}
\href{https://doi.org/10.1109/ICSE.2015.284}{doi:\nolinkurl{10.1109/ICSE.2015.284}}


\bibitem[Musheyev et~al\mbox{.}(2024)]%
        {musheyev2024readability}
\bibfield{author}{\bibinfo{person}{David Musheyev}, \bibinfo{person}{Alexander Pan}, \bibinfo{person}{Preston Gross}, \bibinfo{person}{Daniel Kamyab}, \bibinfo{person}{Peter Kaplinsky}, \bibinfo{person}{Mark Spivak}, \bibinfo{person}{Marie~A Bragg}, \bibinfo{person}{Stacy Loeb}, {and} \bibinfo{person}{Abdo~E Kabarriti}.} \bibinfo{year}{2024}\natexlab{}.
\newblock \showarticletitle{Readability and information quality in cancer information from a free vs paid chatbot}.
\newblock \bibinfo{journal}{\emph{JAMA network open}} \bibinfo{volume}{7}, \bibinfo{number}{7} (\bibinfo{year}{2024}), \bibinfo{pages}{e2422275--e2422275}.
\newblock


\bibitem[Nadi et~al\mbox{.}(2016)]%
        {nadi_jumping_2016}
\bibfield{author}{\bibinfo{person}{Sarah Nadi}, \bibinfo{person}{Stefan Krüger}, \bibinfo{person}{Mira Mezini}, {and} \bibinfo{person}{Eric Bodden}.} \bibinfo{year}{2016}\natexlab{}.
\newblock \showarticletitle{Jumping through hoops: why do Java developers struggle with cryptography {APIs}?}. In \bibinfo{booktitle}{\emph{Proceedings of the 38th International Conference on Software Engineering (ICSE)}} (2016-05-14). \bibinfo{publisher}{{ACM}}, \bibinfo{address}{Austin Texas}, \bibinfo{pages}{935--946}.
\newblock
\showISBNx{978-1-4503-3900-1}
\href{https://doi.org/10.1145/2884781.2884790}{doi:\nolinkurl{10.1145/2884781.2884790}}


\bibitem[Naiakshina et~al\mbox{.}(2020)]%
        {naiakshina_conducting_2020}
\bibfield{author}{\bibinfo{person}{Alena Naiakshina}, \bibinfo{person}{Anastasia Danilova}, \bibinfo{person}{Eva Gerlitz}, {and} \bibinfo{person}{Matthew Smith}.} \bibinfo{year}{2020}\natexlab{}.
\newblock \showarticletitle{On Conducting Security Developer Studies with {CS} Students: Examining a Password-Storage Study with {CS} Students, Freelancers, and Company Developers}. In \bibinfo{booktitle}{\emph{Proceedings of the 2020 {CHI} Conference on Human Factors in Computing Systems}} (2020-04-23) \emph{(\bibinfo{series}{{CHI} '20})}. \bibinfo{publisher}{Association for Computing Machinery}, \bibinfo{address}{New York, {NY}, {USA}}, \bibinfo{pages}{1--13}.
\newblock
\showISBNx{978-1-4503-6708-0}
\href{https://doi.org/10.1145/3313831.3376791}{doi:\nolinkurl{10.1145/3313831.3376791}}


\bibitem[Naiakshina et~al\mbox{.}(2019)]%
        {naiakshina_if_2019}
\bibfield{author}{\bibinfo{person}{Alena Naiakshina}, \bibinfo{person}{Anastasia Danilova}, \bibinfo{person}{Eva Gerlitz}, \bibinfo{person}{Emanuel von Zezschwitz}, {and} \bibinfo{person}{Matthew Smith}.} \bibinfo{year}{2019}\natexlab{}.
\newblock \showarticletitle{"If you want, I can store the encrypted password": A Password-Storage Field Study with Freelance Developers}. In \bibinfo{booktitle}{\emph{Proceedings of the 2019 {CHI} Conference on Human Factors in Computing Systems}} (2019-05-02) \emph{(\bibinfo{series}{{CHI} '19})}. \bibinfo{publisher}{Association for Computing Machinery}, \bibinfo{address}{New York, {NY}, {USA}}, \bibinfo{pages}{1--12}.
\newblock
\showISBNx{978-1-4503-5970-2}
\href{https://doi.org/10.1145/3290605.3300370}{doi:\nolinkurl{10.1145/3290605.3300370}}


\bibitem[Naiakshina et~al\mbox{.}(2017)]%
        {naiakshina_why_2017}
\bibfield{author}{\bibinfo{person}{Alena Naiakshina}, \bibinfo{person}{Anastasia Danilova}, \bibinfo{person}{Christian Tiefenau}, \bibinfo{person}{Marco Herzog}, \bibinfo{person}{Sergej Dechand}, {and} \bibinfo{person}{Matthew Smith}.} \bibinfo{year}{2017}\natexlab{}.
\newblock \showarticletitle{Why Do Developers Get Password Storage Wrong?: A Qualitative Usability Study}. In \bibinfo{booktitle}{\emph{Proceedings of the 2017 {ACM} {SIGSAC} Conference on Computer and Communications Security}} (2017-10-30). \bibinfo{publisher}{{ACM}}, \bibinfo{address}{Dallas Texas {USA}}, \bibinfo{pages}{311--328}.
\newblock
\showISBNx{978-1-4503-4946-8}
\href{https://doi.org/10.1145/3133956.3134082}{doi:\nolinkurl{10.1145/3133956.3134082}}


\bibitem[Naiakshina et~al\mbox{.}(2018)]%
        {naiakshina_deception_2018}
\bibfield{author}{\bibinfo{person}{Alena Naiakshina}, \bibinfo{person}{Anastasia Danilova}, \bibinfo{person}{Christian Tiefenau}, {and} \bibinfo{person}{Matthew Smith}.} \bibinfo{year}{2018}\natexlab{}.
\newblock \showarticletitle{Deception task design in developer password studies: exploring a student sample}. In \bibinfo{booktitle}{\emph{Proceedings of the Fourteenth {USENIX} Conference on Usable Privacy and Security}} (2018) \emph{(\bibinfo{series}{{SOUPS} '18})}. \bibinfo{publisher}{{USENIX} Association}, \bibinfo{address}{{USA}}, \bibinfo{pages}{297--313}.
\newblock
\showISBNx{978-1-931971-45-4}
\newblock
\shownote{event-place: Baltimore, {MD}, {USA}}.


\bibitem[Ng et~al\mbox{.}(2024)]%
        {ng2024harnessing}
\bibfield{author}{\bibinfo{person}{Kevin~KB Ng}, \bibinfo{person}{Liyana Fauzi}, \bibinfo{person}{Leon Leow}, {and} \bibinfo{person}{Jaren Ng}.} \bibinfo{year}{2024}\natexlab{}.
\newblock \bibinfo{title}{Harnessing the Potential of Gen-AI Coding Assistants in Public Sector Software Development}.
\newblock


\bibitem[of~Standards and Technology(2024)]%
        {Cybersec32:online}
\bibfield{author}{\bibinfo{person}{National~Institute of Standards} {and} \bibinfo{person}{Technology}.} \bibinfo{year}{2024}\natexlab{}.
\newblock \bibinfo{title}{Cybersecurity and Privacy}.
\newblock \bibinfo{howpublished}{\url{https://www.nist.gov/cybersecurity-and-privacy}}.
\newblock
\newblock
\shownote{[Online; accessed 2025-12-05]}.


\bibitem[Oh et~al\mbox{.}(2024)]%
        {oh_poisoned_2023}
\bibfield{author}{\bibinfo{person}{Sanghak Oh}, \bibinfo{person}{Kiho Lee}, \bibinfo{person}{Seonhye Park}, \bibinfo{person}{Doowon Kim}, {and} \bibinfo{person}{Hyoungshick Kim}.} \bibinfo{year}{2024}\natexlab{}.
\newblock \showarticletitle{{ Poisoned ChatGPT Finds Work for Idle Hands: Exploring Developers’ Coding Practices with Insecure Suggestions from Poisoned AI Models }}. In \bibinfo{booktitle}{\emph{2024 IEEE Symposium on Security and Privacy (SP)}}. \bibinfo{publisher}{IEEE Computer Society}, \bibinfo{address}{Los Alamitos, CA, USA}, \bibinfo{pages}{1141--1159}.
\newblock
\href{https://doi.org/10.1109/SP54263.2024.00046}{doi:\nolinkurl{10.1109/SP54263.2024.00046}}


\bibitem[OpenAI(2025)]%
        {ChatGPTO21:online}
\bibfield{author}{\bibinfo{person}{OpenAI}.} \bibinfo{year}{2025}\natexlab{}.
\newblock \bibinfo{title}{ChatGPT | OpenAI}.
\newblock \bibinfo{howpublished}{\url{https://chatgpt.com/}}.
\newblock
\newblock
\shownote{[Online; accessed 2025-12-05]}.


\bibitem[Overflow(2023)]%
        {StackOve10:online}
\bibfield{author}{\bibinfo{person}{Stack Overflow}.} \bibinfo{year}{2023}\natexlab{}.
\newblock \bibinfo{title}{Stack Overflow Developer Survey 2023}.
\newblock \bibinfo{howpublished}{\url{https://survey.stackoverflow.co/2023}}.
\newblock
\newblock
\shownote{[Online; accessed 2025-12-05]}.


\bibitem[OWASP(2025)]%
        {A02Crypt35:online}
\bibfield{author}{\bibinfo{person}{OWASP}.} \bibinfo{year}{2025}\natexlab{}.
\newblock \bibinfo{title}{A02 Cryptographic Failures - OWASP Top 10:2021}.
\newblock
\urldef\tempurl%
\url{https://owasp.org/Top10/2021/A02_2021-Cryptographic_Failures/}
\showURL{%
\tempurl}
\newblock
\shownote{[Online; accessed 2025-12-10]}.


\bibitem[OWASP~Foundation(2024a)]%
        {Aboutthe44:online}
\bibfield{author}{\bibinfo{person}{Inc. OWASP~Foundation}.} \bibinfo{year}{2024}\natexlab{a}.
\newblock \bibinfo{title}{About the OWASP Foundation | OWASP Foundation}.
\newblock \bibinfo{howpublished}{\url{https://owasp.org/about/}}.
\newblock
\newblock
\shownote{[Online; accessed 2025-12-05]}.


\bibitem[OWASP~Foundation(2024b)]%
        {SourceCo79:online}
\bibfield{author}{\bibinfo{person}{Inc. OWASP~Foundation}.} \bibinfo{year}{2024}\natexlab{b}.
\newblock \bibinfo{title}{Source Code Analysis Tools | OWASP Foundation}.
\newblock \bibinfo{howpublished}{\url{https://owasp.org/www-community/Source\_Code\_Analysis\_Tools}}.
\newblock
\newblock
\shownote{[Online; accessed 2025-12-05]}.


\bibitem[Pearce et~al\mbox{.}(2025)]%
        {pearce_asleep_2021}
\bibfield{author}{\bibinfo{person}{Hammond Pearce}, \bibinfo{person}{Baleegh Ahmad}, \bibinfo{person}{Benjamin Tan}, \bibinfo{person}{Brendan Dolan-Gavitt}, {and} \bibinfo{person}{Ramesh Karri}.} \bibinfo{year}{2025}\natexlab{}.
\newblock \showarticletitle{Asleep at the Keyboard? Assessing the Security of GitHub Copilot’s Code Contributions}.
\newblock \bibinfo{journal}{\emph{Commun. ACM}} \bibinfo{volume}{68}, \bibinfo{number}{2} (\bibinfo{date}{Jan.} \bibinfo{year}{2025}), \bibinfo{pages}{96–105}.
\newblock
\showISSN{0001-0782}
\href{https://doi.org/10.1145/3610721}{doi:\nolinkurl{10.1145/3610721}}


\bibitem[Pearce et~al\mbox{.}(2023)]%
        {pearce_examining_2023}
\bibfield{author}{\bibinfo{person}{Hammond Pearce}, \bibinfo{person}{Benjamin Tan}, \bibinfo{person}{Baleegh Ahmad}, \bibinfo{person}{Ramesh Karri}, {and} \bibinfo{person}{Brendan Dolan-Gavitt}.} \bibinfo{year}{2023}\natexlab{}.
\newblock \showarticletitle{Examining Zero-Shot Vulnerability Repair with Large Language Models}. In \bibinfo{booktitle}{\emph{2023 {IEEE} Symposium on Security and Privacy ({SP})}} (2023-05). \bibinfo{publisher}{{IEEE}}, \bibinfo{address}{San Francisco, {CA}, {USA}}, \bibinfo{pages}{2339--2356}.
\newblock
\showISBNx{978-1-66549-336-9}
\href{https://doi.org/10.1109/SP46215.2023.10179324}{doi:\nolinkurl{10.1109/SP46215.2023.10179324}}


\bibitem[Percival(2009)]%
        {percival_stronger_2009}
\bibfield{author}{\bibinfo{person}{Colin Percival}.} \bibinfo{year}{2009}\natexlab{}.
\newblock \bibinfo{title}{Stronger key derivation via sequential memory-hard functions}.
\newblock


\bibitem[Perry et~al\mbox{.}(2023)]%
        {perry_users_2022}
\bibfield{author}{\bibinfo{person}{Neil Perry}, \bibinfo{person}{Megha Srivastava}, \bibinfo{person}{Deepak Kumar}, {and} \bibinfo{person}{Dan Boneh}.} \bibinfo{year}{2023}\natexlab{}.
\newblock \showarticletitle{Do Users Write More Insecure Code with AI Assistants?}. In \bibinfo{booktitle}{\emph{Proceedings of the 2023 ACM SIGSAC Conference on Computer and Communications Security}} (Copenhagen, Denmark,) \emph{(\bibinfo{series}{CCS '23})}. \bibinfo{publisher}{Association for Computing Machinery}, \bibinfo{address}{New York, NY, USA}, \bibinfo{pages}{2785–2799}.
\newblock
\showISBNx{9798400700507}
\href{https://doi.org/10.1145/3576915.3623157}{doi:\nolinkurl{10.1145/3576915.3623157}}


\bibitem[Petkauskas(2024)]%
        {cybernews}
\bibfield{author}{\bibinfo{person}{Vilius Petkauskas}.} \bibinfo{year}{2024}\natexlab{}.
\newblock \bibinfo{title}{RockYou2024: 10 billion passwords leaked in the largest compilation of all time}.
\newblock \bibinfo{howpublished}{\url{https://cybernews.com/security/rockyou2024-largest-password-compilation-leak/}}.
\newblock
\newblock
\shownote{[Online; accessed 2025-12-05]}.


\bibitem[Pichai(2024)]%
        {GoogleGe4:online}
\bibfield{author}{\bibinfo{person}{Sundar Pichai}.} \bibinfo{year}{2024}\natexlab{}.
\newblock \bibinfo{title}{Google Gemini update: Sundar Pichai introduces Ultra 1.0 in Gemini Advanced}.
\newblock
\urldef\tempurl%
\url{https://blog.google/technology/ai/google-gemini-update-sundar-pichai-2024/}
\showURL{%
\tempurl}
\newblock
\shownote{[Online; accessed 2025-05-13]}.


\bibitem[{Prolific}(2025)]%
        {prolific_llm_detection}
\bibfield{author}{\bibinfo{person}{{Prolific}}.} \bibinfo{year}{2025}\natexlab{}.
\newblock \bibinfo{title}{How to Detect and Prevent the Use of Large Language Models in Studies}.
\newblock \bibinfo{howpublished}{\url{https://researcher-help.prolific.com/en/articles/445207-how-to-detect-and-prevent-the-use-of-large-language-models-in-studies}}.
\newblock
\newblock
\shownote{Prolific Research Help Center. Accessed: 2026-03-13}.


\bibitem[Provos and Mazi\`{e}res(1999)]%
        {provos_future-adaptable_1997}
\bibfield{author}{\bibinfo{person}{Niels Provos} {and} \bibinfo{person}{David Mazi\`{e}res}.} \bibinfo{year}{1999}\natexlab{}.
\newblock \showarticletitle{A future-adaptive password scheme}. In \bibinfo{booktitle}{\emph{Proceedings of the Annual Conference on USENIX Annual Technical Conference}} (Monterey, California) \emph{(\bibinfo{series}{ATEC '99})}. \bibinfo{publisher}{USENIX Association}, \bibinfo{address}{USA}, \bibinfo{pages}{32}.
\newblock


\bibitem[(PyCQA)(2024)]%
        {PyCQAban31:online}
\bibfield{author}{\bibinfo{person}{Python Code Quality~Authority (PyCQA)}.} \bibinfo{year}{2024}\natexlab{}.
\newblock \bibinfo{title}{PyCQA/bandit: Bandit is a tool designed to find common security issues in Python code.}
\newblock \bibinfo{howpublished}{\url{https://github.com/PyCQA/bandit}}.
\newblock
\newblock
\shownote{[Online; accessed 2025-12-05]}.


\bibitem[SA.(2024)]%
        {CodeQual5:online}
\bibfield{author}{\bibinfo{person}{SonarSource SA.}} \bibinfo{year}{2024}\natexlab{}.
\newblock \bibinfo{title}{Code Quality, Security \& Static Analysis Tool with SonarQube | Sonar}.
\newblock \bibinfo{howpublished}{\url{https://www.sonarsource.com/products/sonarqube/}}.
\newblock
\newblock
\shownote{[Online; accessed 2025-12-05]}.


\bibitem[Sandoval et~al\mbox{.}(2023)]%
        {sandoval_lost_2022}
\bibfield{author}{\bibinfo{person}{Gustavo Sandoval}, \bibinfo{person}{Hammond Pearce}, \bibinfo{person}{Teo Nys}, \bibinfo{person}{Ramesh Karri}, \bibinfo{person}{Siddharth Garg}, {and} \bibinfo{person}{Brendan Dolan-Gavitt}.} \bibinfo{year}{2023}\natexlab{}.
\newblock \showarticletitle{Lost at C: A User Study on the Security Implications of Large Language Model Code Assistants}. In \bibinfo{booktitle}{\emph{32nd USENIX Security Symposium (USENIX Security 23)}}. \bibinfo{publisher}{USENIX Association}, \bibinfo{address}{Anaheim, CA}, \bibinfo{pages}{2205--2222}.
\newblock
\showISBNx{978-1-939133-37-3}
\urldef\tempurl%
\url{https://www.usenix.org/conference/usenixsecurity23/presentation/sandoval}
\showURL{%
\tempurl}


\bibitem[Siddiq et~al\mbox{.}(2022)]%
        {siddiq_empirical_2022}
\bibfield{author}{\bibinfo{person}{Mohammed~Latif Siddiq}, \bibinfo{person}{Shafayat~H. Majumder}, \bibinfo{person}{Maisha~R. Mim}, \bibinfo{person}{Sourov Jajodia}, {and} \bibinfo{person}{Joanna C.~S. Santos}.} \bibinfo{year}{2022}\natexlab{}.
\newblock \showarticletitle{An Empirical Study of Code Smells in Transformer-based Code Generation Techniques}. In \bibinfo{booktitle}{\emph{2022 {IEEE} 22nd International Working Conference on Source Code Analysis and Manipulation ({SCAM})}} (2022-10). \bibinfo{publisher}{{IEEE}}, \bibinfo{address}{Limassol, Cyprus}, \bibinfo{pages}{71--82}.
\newblock
\showISBNx{978-1-66549-609-4}
\href{https://doi.org/10.1109/SCAM55253.2022.00014}{doi:\nolinkurl{10.1109/SCAM55253.2022.00014}}


\bibitem[Smith(2024)]%
        {CanGener5:online}
\bibfield{author}{\bibinfo{person}{Andy Smith}.} \bibinfo{year}{2024}\natexlab{}.
\newblock \bibinfo{title}{Can Generative AI Solve the Software Engineer Shortage? | HatchWorks AI}.
\newblock
\urldef\tempurl%
\url{https://hatchworks.com/blog/gen-ai/software-engineer-shortage/}
\showURL{%
\tempurl}
\newblock
\shownote{[Online; accessed 2025-12-05]}.


\bibitem[Sobania et~al\mbox{.}(2022)]%
        {sobania_choose_2022}
\bibfield{author}{\bibinfo{person}{Dominik Sobania}, \bibinfo{person}{Martin Briesch}, {and} \bibinfo{person}{Franz Rothlauf}.} \bibinfo{year}{2022}\natexlab{}.
\newblock \showarticletitle{Choose your programming copilot: a comparison of the program synthesis performance of github copilot and genetic programming}. In \bibinfo{booktitle}{\emph{Proceedings of the Genetic and Evolutionary Computation Conference}} (2022-07-08) \emph{(\bibinfo{series}{{GECCO} '22})}. \bibinfo{publisher}{Association for Computing Machinery}, \bibinfo{address}{New York, {NY}, {USA}}, \bibinfo{pages}{1019--1027}.
\newblock
\showISBNx{978-1-4503-9237-2}
\href{https://doi.org/10.1145/3512290.3528700}{doi:\nolinkurl{10.1145/3512290.3528700}}


\bibitem[{Stack Exchange, Inc.}(2022)]%
        {StackOve32:online}
\bibfield{author}{\bibinfo{person}{{Stack Exchange, Inc.}}} \bibinfo{year}{2022}\natexlab{}.
\newblock \bibinfo{title}{Stack Overflow Developer Survey 2022}.
\newblock \bibinfo{howpublished}{\url{https://survey.stackoverflow.co/2022/##demographics-gender}}.
\newblock
\newblock
\shownote{[Online; accessed 2025-12-05]}.


\bibitem[{Stack Exchange, Inc.}(2024)]%
        {2024Stac15:online}
\bibfield{author}{\bibinfo{person}{{Stack Exchange, Inc.}}} \bibinfo{year}{2024}\natexlab{}.
\newblock \bibinfo{title}{2024 Stack Overflow Developer Survey}.
\newblock \bibinfo{howpublished}{\url{https://survey.stackoverflow.co/2024/}}.
\newblock
\newblock
\shownote{[Online; accessed 2025-12-05]}.


\bibitem[Statista(2022a)]%
        {Authenti35:online}
\bibfield{author}{\bibinfo{person}{Statista}.} \bibinfo{year}{2022}\natexlab{a}.
\newblock \bibinfo{title}{Authentication factors used by finance companies 2022 | Statista}.
\newblock \bibinfo{howpublished}{\url{https://www.statista.com/statistics/1342535/authentication-factors-used-by-finance-companies/}}.
\newblock
\newblock
\shownote{[Online; accessed 2025-12-05]}.


\bibitem[Statista(2022b)]%
        {Authenti52:online}
\bibfield{author}{\bibinfo{person}{Statista}.} \bibinfo{year}{2022}\natexlab{b}.
\newblock \bibinfo{title}{Authentication factors used by healthcare companies 2022 | Statista}.
\newblock \bibinfo{howpublished}{\url{https://www.statista.com/statistics/1342533/authentication-factors-used-by-healthcare-companies/}}.
\newblock
\newblock
\shownote{[Online; accessed 2025-12-05]}.


\bibitem[Statista(2024)]%
        {cloud_revenue}
\bibfield{author}{\bibinfo{person}{Statista}.} \bibinfo{year}{2024}\natexlab{}.
\newblock \bibinfo{title}{Revenue of the public cloud market worldwide from 2020 to 2029}.
\newblock \bibinfo{howpublished}{\url{https://www.statista.com/forecasts/963841/cloud-services-revenue-in-the-world}}.
\newblock
\newblock
\shownote{[Online; accessed 2025-12-05]}.


\bibitem[Tahaei et~al\mbox{.}(2021)]%
        {tahaei_security_2021}
\bibfield{author}{\bibinfo{person}{Mohammad Tahaei}, \bibinfo{person}{Kami Vaniea}, \bibinfo{person}{Konstantin~(Kosta) Beznosov}, {and} \bibinfo{person}{Maria~K Wolters}.} \bibinfo{year}{2021}\natexlab{}.
\newblock \showarticletitle{Security Notifications in Static Analysis Tools: Developers’ Attitudes, Comprehension, and Ability to Act on Them}. In \bibinfo{booktitle}{\emph{Proceedings of the 2021 {CHI} Conference on Human Factors in Computing Systems}} (2021-05-07) \emph{(\bibinfo{series}{{CHI} '21})}. \bibinfo{publisher}{Association for Computing Machinery}, \bibinfo{address}{New York, {NY}, {USA}}, \bibinfo{pages}{1--17}.
\newblock
\showISBNx{978-1-4503-8096-6}
\href{https://doi.org/10.1145/3411764.3445616}{doi:\nolinkurl{10.1145/3411764.3445616}}


\bibitem[{The MITRE Corporation}(2024a)]%
        {CWE2023C20:online}
\bibfield{author}{\bibinfo{person}{{The MITRE Corporation}}.} \bibinfo{year}{2024}\natexlab{a}.
\newblock \bibinfo{title}{CWE - 2023 CWE Top 25 Most Dangerous Software Weaknesses}.
\newblock \bibinfo{howpublished}{\url{{https://cwe.mitre.org/top25/archive/2023/2023\_top25\_list.html}}}.
\newblock
\newblock
\shownote{[Online; accessed 2025-12-05]}.


\bibitem[{The MITRE Corporation}(2024b)]%
        {CWECWE2047:online}
\bibfield{author}{\bibinfo{person}{{The MITRE Corporation}}.} \bibinfo{year}{2024}\natexlab{b}.
\newblock \bibinfo{title}{CWE - CWE-20: Improper Input Validation (4.15)}.
\newblock \bibinfo{howpublished}{\url{https://cwe.mitre.org/data/definitions/20.html}}.
\newblock
\newblock
\shownote{[Online; accessed 2025-12-05]}.


\bibitem[{The MITRE Corporation}(2024c)]%
        {CWECWE3536:online}
\bibfield{author}{\bibinfo{person}{{The MITRE Corporation}}.} \bibinfo{year}{2024}\natexlab{c}.
\newblock \bibinfo{title}{CWE - CWE-352: Cross-Site Request Forgery (CSRF) (4.15)}.
\newblock \bibinfo{howpublished}{\url{https://cwe.mitre.org/data/definitions/352.html}}.
\newblock
\newblock
\shownote{[Online; accessed 2025-12-05]}.


\bibitem[{The MITRE Corporation}(2024d)]%
        {CWECWE792:online}
\bibfield{author}{\bibinfo{person}{{The MITRE Corporation}}.} \bibinfo{year}{2024}\natexlab{d}.
\newblock \bibinfo{title}{CWE - CWE-79: Improper Neutralization of Input During Web Page Generation ('Cross-site Scripting') (4.15)}.
\newblock \bibinfo{howpublished}{\url{https://cwe.mitre.org/data/definitions/79.html}}.
\newblock
\newblock
\shownote{[Online; accessed 2025-12-05]}.


\bibitem[{The MITRE Corporation}(2024e)]%
        {CWECWE8999:online}
\bibfield{author}{\bibinfo{person}{{The MITRE Corporation}}.} \bibinfo{year}{2024}\natexlab{e}.
\newblock \bibinfo{title}{CWE - CWE-89: Improper Neutralization of Special Elements used in an SQL Command ('SQL Injection') (4.15)}.
\newblock \bibinfo{howpublished}{\url{https://cwe.mitre.org/data/definitions/89.html}}.
\newblock
\newblock
\shownote{[Online; accessed 2025-12-05]}.


\bibitem[Votipka et~al\mbox{.}(2020)]%
        {votipka_building_2020}
\bibfield{author}{\bibinfo{person}{Daniel Votipka}, \bibinfo{person}{Desiree Abrokwa}, {and} \bibinfo{person}{Michelle~L. Mazurek}.} \bibinfo{year}{2020}\natexlab{}.
\newblock \showarticletitle{Building and Validating a Scale for Secure Software Development Self-Efficacy}. In \bibinfo{booktitle}{\emph{Proceedings of the 2020 {CHI} Conference on Human Factors in Computing Systems}} (Honolulu {HI} {USA}, 2020-04-21). \bibinfo{publisher}{{ACM}}, \bibinfo{address}{Honolulu, HI, USA}, \bibinfo{pages}{1--20}.
\newblock
\showISBNx{978-1-4503-6708-0}
\href{https://doi.org/10.1145/3313831.3376754}{doi:\nolinkurl{10.1145/3313831.3376754}}


\bibitem[Zapponi(2024)]%
        {GithubLa92:online}
\bibfield{author}{\bibinfo{person}{Carlo Zapponi}.} \bibinfo{year}{2024}\natexlab{}.
\newblock \bibinfo{title}{Github Language Stats}.
\newblock \bibinfo{howpublished}{\url{https://madnight.github.io/githut/}}.
\newblock
\newblock
\shownote{[Online; accessed 2025-12-05]}.


\end{thebibliography}

\appendix

\section{Appendix}\label{sec:appendix}

\subsection{Availability}
Within %
the appendix, we outline the following components:
\begin{enumerate}
    \item \textbf{Section 2:} Within this section, we provide the \textbf{job post} used for recruiting developers on Upwork.
    \item \textbf{Section 3:} This section details the \textbf{task description} presented to the participants.
    \item \textbf{Section 4:} Here, we present the follow-up \textbf{survey}.
    \item \textbf{Section 5:} This section provides our \textbf{codebook}.
\end{enumerate}
Participants’ solutions and responses are not provided in order to protect their privacy.

\subsection{Job Post}\label{jobpost}
\textbf{Participants for Python Programming Study with Flask}\\
We are researchers from [anonymous university] and are looking for software developers who are interested in participating in our programming study.

Our research goal is to study developers' programming behavior with a Flask web application. You will be asked to complete four short programming tasks in a Python Flask web application, submit the code via GitHub Classroom, and fill out a short survey.

You will be rewarded with an expense allowance of \$60. The study will be conducted in English. You are free to organize your time as you wish, but the tasks must be completed within 2 days (48h) after accepting the job.

All data will be processed pseudonymously and stored anonymized after the study; there will be no identifying information published in any form.

What does this job look like?
\begin{enumerate}
    \item Apply for the job
    \item Chose a preferred appointment
    \item Complete the programming task
    \item Fill out a short survey
    \item Receive payment via Upwork
\end{enumerate}

Please complete the study in a focused setting, using a laptop or PC.

\subsection{Task Description}\label{task}

Welcome! We would like to invite you to participate in our study and thus make a valuable contribution to our research. Our research goal is to study developers' programming behavior with a Flask web application.

Instructions:
\begin{itemize}
    \item The task requires Python (at least version 3.10.)
    \item You can use an IDE of your choice for solving the task. Before starting with the task, please disable all IDE Plugins that use AI (plugins such as GitHub Copilot, Tabnine, …)
    \item \textbf{(\No{} group:)} Please do not use any kind of AI to fulfill this task\\
    \textbf{(\Free{} group:)} Please only use the free version of Gemini (https://gemini.google.com) for solving the task. If you do not have a Google account, please create one for free.
    \textbf{(\Paid{} group:)} Please only use the paid version of Gemini (https://gemini.google.com) for solving the task. The access data for a Google account with a paid Google One subscription (AI Premium Version) will be provided. Within this account you will have access to Gemini Advanced.
    \begin{enumerate}
        \item Follow the link: https://accounts.google.com/ 
        \item Login with the provided Google Account:
        \item Username: (username)
        \item Password: (password)
        \item Follow the link https://gemini.google.com to use Gemini
    \end{enumerate}
    \item \textbf{(\Paid{} group:)} Please do not enter login data such as passwords or personal data into the chat in Gemini. All chats in the provided account will be deleted after the participation.
    \item \textbf{(\Free{} group:)} Please do not use any other AI tools except the free version of Gemini.
    \item \textbf{(\Paid{} group:)} Please do not use any other AI tools except the paid version of Gemini.
    \item In order to solve the task, you are allowed to use any kind of source available on the Internet that may be helpful or has valuable information, except Websites that include AI such as ChatGPT, Microsoft Copilot/Bing Chat \textbf{(\No{} group:)} or Google Gemini.\\
    \textbf{(\Free{} group:)} Please use only the free version of Gemini instead.\\
    \textbf{(\Paid{} group:)} Please use only the premium version of Gemini instead.
    \item You may import any needed external libraries.
    \item How to start the application: python -m flask run
\end{itemize}

Link to Task: [Link to GitHub Classroom]

Task:
\begin{itemize}
    \item Please follow the link above and connect your GitHub Account to GitHub Classroom. Select your Pseudonym from the list.
    \item Clone the code to your IDE
    \item Install the requirements with pip install -r requirements.txt
    \item Complete the TODOs in the code
    \item Push the completed code to GitHub
    \item Please fill in the following survey after working on the task: [Link to survey on Qualtrics]
\end{itemize}

Implementation Hints:\\
The Python code sets up a Flask web application with user authentication and basic CRUD functionalities for managing user websites. It uses Flask's login extension for managing user sessions and SQLite for database operations. The User class represents user objects with their credentials. There are routes for registering, logging in, logging out, viewing the dashboard and managing websites. Functionalities that should be implemented are marked with TODO. Please pay attention to the security aspects during implementation. In the future, it is planned that users will be able to share their profiles containing their websites publicly, therefore making them visible to others.

\subsection{Survey}\label{survey}
\begin{itemize}
    \item Please enter your pseudonym:
    \item How old are you?
    \item What is your gender?
    \begin{itemize}
        \item Woman
        \item Man
        \item Non-binary
        \item Prefer not to disclose
        \item Prefer to self-describe:
    \end{itemize}
    \item In which country do you currently live mainly?
    \item What level of education do you have? Please indicate only the highest degree.
    \begin{itemize}
        \item No school degree
        \item School degree
        \item Professional training
        \item Bachelor Degree
        \item Masters Degree or equivalent Diploma
        \item Doctoral degree (Dr./PhD)
        \item Other university degree
        \item Other, namely:
    \end{itemize}
    \item Please specify the number of years you have been programming in general:
    \item Please specify the number of years you have been programming in Python:
    \item Which programming, scripting, and markup language(s) do you regularly utilize?
    \begin{itemize}
        \item Bash/Shell (all shells)
        \item C
        \item C\#
        \item C++
        \item HTML/CSS
        \item Java
        \item JavaScript
        \item Python
        \item SQL
        \item TypeScript
        \item Other:
    \end{itemize}
    \item Which Integrated Development Environment (IDE) do you commonly use?
    \begin{itemize}
        \item Android Studio
        \item IntelliJ IDEA
        \item Jupyter Notebook/JupyterLab
        \item Neovim
        \item Notepad++
        \item PyCharm
        \item Sublime Text
        \item Vim
        \item Visual Studio
        \item Visual Studio Code
        \item Other:
    \end{itemize}
    \item Which IDE did you utilize for completing the task?
    \begin{itemize}
        \item Android Studio
        \item IntelliJ IDEA
        \item Jupyter Notebook/JupyterLab
        \item Neovim
        \item Notepad++
        \item PyCharm
        \item Sublime Text
        \item Vim
        \item Visual Studio
        \item Visual Studio Code
        \item Other:
    \end{itemize}
    \item Which AI tools do you commonly use?
    \begin{itemize}
        \item None
        \item ChatGPT Free Version
        \item ChatGPT Paid Version
        \item Gemini Free Version
        \item Gemini Paid Version
        \item GitHub Copilot
        \item Tabnine Free Version
        \item Tabnine Paid Version
        \item Visual Studio IntelliCode
        \item Other:
    \end{itemize}
    \item Which AI tools did you utilize for solving the task?
    \begin{itemize}
        \item None
        \item ChatGPT Free Version
        \item ChatGPT Paid Version
        \item Gemini Free Version
        \item Gemini Paid Version
        \item GitHub Copilot
        \item Tabnine Free Version
        \item Tabnine Paid Version
        \item Visual Studio IntelliCode
        \item Other:
    \end{itemize}
    \item Do you have IT security experience?
    \begin{itemize}
        \item No experience
        \item I have taken an IT security course/training
        \item I have a degree in IT security
        \item I have a certificate in IT security
        \item I have developed security applications/implemented security measures
        \item I have worked at IT security-related companies
        \item I have worked on IT security in my spare time
        \item Other:
    \end{itemize}
    \item What is your current main occupation?
    \begin{itemize}
        \item Freelance developer
        \item Industrial developer
        \item Freelance security expert
        \item Industrial security expert
        \item Industrial researcher
        \item Academic researcher
        \item Undergraduate student
        \item Graduate student
        \item Other:
    \end{itemize}
    \item How much time in minutes did you need to solve the task?
    \item On a scale of 1 to 7, please rate the following statement: 'I believe I have solved this task functionally correctly.' \\
    1: Strongly Disagree - 7: Strongly Agree
    \item On a scale of 1 to 7, how would you rate the level of security of your solution for the task? \\
    1: Not Secure - 7: Totally Secure
    \item \textbf{(\Free{} \& \Paid{} group:)} 
    \begin{itemize}
        \item On a scale from 1 to 7, how helpful do you consider Gemini in solving this task?\\
        1: Not Helpful - 7: Very Helpful
    \item In what ways has the assistance provided by Gemini positively influenced the solving of the task?
    \item In what ways has the assistance provided by Gemini presented challenges in solving the task?
    \item On a scale of 1 to 7, please rate the following statement: 'I trust Gemini in general.'\\
    1: Never - 7: Always
    \item On a scale of 1 to 7, please rate the following statement: 'I trust Gemini to generate secure code.'\\
    1: Never - 7: Always
    \item Why did you choose to trust the suggestions by Gemini?
    \item Why did you choose to not trust the suggestions by Gemini?
    \item To help us understand your interactions with Gemini, please answer the following questions honestly:
    \begin{itemize}
        \item How often have you consulted Gemini (questions or prompts)?
        \item What percentage of Gemini's suggestions did you adopt?\\
        0\% - 100 \%
    \end{itemize}
    \end{itemize}
    \item \textbf{(\No{} group:)} 
    \begin{itemize}
        \item Would you have appreciated assistance from an AI?
        \begin{itemize}
            \item Yes
            \item No
        \end{itemize}
        \item On a scale of 1 to 7, would you trust AI in general? \\
        1: Never - 7: Always
        \item On a scale of 1 to 7, would you trust AI to generate secure code? \\
        1: Never - 7: Always
    \end{itemize}
    \item Can you describe any particular challenges you faced during the development process?
    \item \textbf{(\No{} group:)} Which resources did you use to complete the task?\\
    \textbf{(\Free{} \& \Paid{} group:)} Which resources (beside Gemini) did you use to complete the task?
    \begin{itemize}
        \item Codecademy
        \item GitHub
        \item Official documentation
        \item StackOverflow
        \item Tutorial Websites
        \item W3Schools
        \item YouTube
        \item Other:
    \end{itemize}
    \item Were there any industry best practices or standards that you followed to enhance security in your development process? If yes, please name them.
    \begin{itemize}
        \item No
        \item Yes:
    \end{itemize}
    \item NASA-TLX \cite{hart1988development} 

    \item Secure Software Development Self-Efficacy Scale (SSD-SES) \cite{votipka_building_2020} 

    \item \textbf{(\Free{} \& \Paid{} group:)} 
    System Usability Scale (SUS) \cite{brooke1996sus} 
\end{itemize}

\subsection{Codebook}\label{codebook}
\begin{itemize}
    \item Codesystem: 1327
    \item Other AI tools: 0
    \begin{itemize}
        \item ChatGPT: 4
        \item Copilot: 2
        \item BlackBoxAI: 1
    \end{itemize}
    
    \item Gemini challenges: 0
    \begin{itemize}
        \item Quality of suggestions: 0
        \begin{itemize}
            \item Suggestions not working: 36
            \item Solution not optimal: 19
            \item Incomplete suggestions: 18
            \item Output needs to be tested and adjusted: 12
            \item Inaccurate code: 10
            \item Vague answers: 10
            \item Unuseful suggestions: 8
            \item Long answers: 5
            \item Outdated suggestions: 4
            \item Inconsistency: 4
            \item Hallucinations: 4
            \item Not mentioning needed library: 4
            \item Not for complex problems: 3
            \item Code formatting/indentation: 2
        \end{itemize}
        \item Usability: 0
        \begin{itemize}
            \item Difficulties prompting: 7
            \item Copy\&Pasting/Non-printable characters in output: 4
            \item Hard to convince tool to do other approach: 4
            \item UI: 4
            \item Integrating suggestions into existing code: 3
            \item Slow: 2
            \item Hard to pass files to Gemini: 1
        \end{itemize}
        \item Missing context: 23
        \item Security: 21
        \item Multiple prompts/re-prompting: 17
        \item No challenges: 12
        \item Misunderstanding of problem/requirements: 9
        \item Suggestions for Gemini improvement: 4
        \item LLM can't assist message: 3
        \item Personal preferences: 2
        \item Gemini gives results from Google, no new code: 2
        \item Lack of experience with Gemini: 1
        \item Less own understanding of code: 1
    \end{itemize}    
    \item Gemini trust: 0
    \begin{itemize}
        \item Mistrust: 0
        \begin{itemize}
            \item Assessment: 28
            \item Security flaws: 28
            \item Suggestions not working: 26
            \item Inaccurate: 15
            \item Mistrust in LLM: 8
            \item Forgetting context: 8
            \item Outdated: 7
            \item Training data: 7
            \item Not following/misinterpreting requirements: 7
            \item Suggestions needed modifications: 7
            \item Hallucinations: 5
            \item Vague answers: 5
            \item Less practical suggestions: 4
            \item Inconsistency: 3
            \item Less unique suggestions: 3
            \item Including unnecessary/not optimal libraries: 2
            \item Explanation not convincing: 2
            \item Complex tasks: 2
            \item Lack of own knowledge: 1
            \item New technology: 1
            \item Personal coding preferences: 1
            \item Other AI tools perform better: 1
            \item Gemini disclaimer: 1
        \end{itemize}
        \item Trust: 0
        \begin{itemize}
            \item Assessment: 42
            \item Correct/working suggestions: 17
            \item Explanations/informative suggestions: 13
            \item Security: 10
            \item Training data: 10
            \item Trusted company: 6
            \item Suggestions meet expectations: 5
            \item Simple tasks: 5
            \item References Gemini gave: 4
            \item Well-known: 3
            \item Quick/faster solutions: 3
            \item Only use for initial idea: 3
            \item Lack of own knowledge: 3
            \item Past experience with Gemini: 2
            \item Part of task: 2
            \item It's the best: 2
            \item Trust in LLM: 2
            \item Other AI tools perform worse: 2
            \item Premium version: 2
            \item Presents multiple options: 1
            \item Gemini disclaimer: 1
        \end{itemize}
    \end{itemize}
    
    \item Task challenges: 0
    \begin{itemize}
        \item Security: 7
        \begin{itemize}
            \item Hashing: 16
            \item Unfamiliar flask\_login: 10
            \item Authentication: 6
            \item SQL Injection Prevention: 3
            \item Input validation: 2
            \item Integrating security into existing code: 2
            \item Searching measures for password security: 2
            \item Balancing security and usability: 1
            \item XSS Prevention: 1
        \end{itemize}
        \item No challenges: 36
        \item Database: 24
        \item Flask: 20
        \item Setup: 14
        \item Debugging: 14
        \item Session management: 11
        \item Implement task functionality: 0
        \begin{itemize}
            \item Dashboard view: 4
            \item Website deletion: 2
            \item Delete user: 1
            \item Linking websites to users: 1
            \item Routing Links: 1
        \end{itemize}
        \item SQL Issues: 9
        \item Unfamiliar libraries: 7
        \item Syntax: 6
        \item Task uncertainty: 6
        \item Not using AI: 5
        \item Integration code into existing system: 4
        \item User messages: 4
        \item Understanding code base: 3
        \item Adding modules/libraries: 3
        \item Smaller modules: 2
        \item Time management: 2
        \item Passing parameter to template: 2
        \item Testing: 2
        \item Data storage: 2
        \item Handle variables: 2
        \item Git: 1
        \item Port: 1
        \item Error handling: 1
    \end{itemize}
    
    \item Gemini assistance: 0
    \begin{itemize}
        \item Security: 0
        \begin{itemize}
            \item Security suggestions: 11
            \item Find security issues: 9
            \item Hashing: 8
            \item flask\_login: 4
            \item SQL Injection Prevention: 2
            \item User permissions: 1
        \end{itemize}
        \item Code generation/completion: 37
        \item Templating: 22
        \item Efficiency/Productivity/Saving time: 20
        \item Debugging: 20
        \item Guidance: 18
        \item Understand codebase: 17
        \item Database: 13
        \item Flask: 9
        \item Syntax: 8
        \item For simple tasks/basic code: 7
        \item Evaluate possible solutions: 6
        \item Pytest/Testing: 4
        \item Setup: 4
        \item Recommend libraries: 3
        \item Improved accuracy: 3
        \item Integrate libraries: 3
        \item Git: 3
        \item Error handling: 2
    \end{itemize}
    
    \item Security best practices: 0
    \begin{itemize}
        \item Hashing: 47
        \begin{itemize}
            \item BCrypt: 11
            \item werkzeug.security: 4
            \item SECRETS library: 1
            \item argon2: 1
            \item SHA-256: 1
        \end{itemize}
        \item SQL injection prevention: 19
        \item Authentication: 17
        \item Input validation: 9
        \item Error handling: 8
        \item Encryption: 7
        \item Library: 7
        \item Misconceptions: 6
        \item Managing sessions: 5
        \item Flak security: 3
        \begin{itemize}
            \item Flask login: 2
        \end{itemize}
        \item CSFR protection: 4
        \item Sanitizing: 4
        \item PEP8: 4
        \item Password requirements: 4
        \item OWASP: 1
        \begin{itemize}
            \item OWASP Top 10: 1
            \item OWASP Secure Coding: 1
            \item OWASP Cheat sheet: 1
        \end{itemize}
        \item Database security: 3
        \item Functionality first, security second: 2
        \item Needs more security information: 2
        \item XSS Prevention: 2
        \item GDPR: 1
        \item NIST: 1
        \item Used tested libraries: 1
        \item Logging: 1
    \end{itemize}
\end{itemize}

\end{document}